\documentclass{aastex63}
\received{March 26, 2022}
\revised{May 15, 2022}
\accepted{May 17, 2022}
\submitjournal{ApJS}

\shorttitle{2D compare}
\shortauthors{Yoo et al.}
\graphicspath{{./}{figures/}}

\begin{document}

\title{
Comparison of spatial distributions of Intracluster light and Dark Matter}

\author[0000-0002-6841-8329]{Jaewon Yoo}
\affil{Korea Astronomy and Space Science Institute (KASI),
Daedeokdae-ro, Daejeon 34055, Korea}
\affil{University of Science and Technology (UST),
Gajeong-ro, Daejeon 34113, Korea}

\author[0000-0002-9434-5936]{Jongwan Ko}
\affil{Korea Astronomy and Space Science Institute (KASI),
Daedeokdae-ro, Daejeon 34055, Korea}
\affil{University of Science and Technology (UST),
Gajeong-ro, Daejeon 34113, Korea}

\author[0000-0002-5513-5303]{Cristiano G. Sabiu}
\affil{Natural Science Research Institute (NSRI), University of Seoul, Seoul 02504, Korea}

\author[0000-0001-5135-1693]{Jihye Shin}
\affil{Korea Astronomy and Space Science Institute (KASI),
Daedeokdae-ro, Daejeon 34055, Korea}

\author[0000-0001-9544-7021]{Kyungwon Chun}
\affil{Korea Astronomy and Space Science Institute (KASI),
Daedeokdae-ro, Daejeon 34055, Korea}

\author[0000-0003-3428-7612]{Ho Seong Hwang}
\affil{Astronomy Program, Department of Physics and Astronomy, Seoul National University (SNU),
1 Gwanak-ro, Gwanak-gu,Seoul 08826, Korea}
\affil{SNU Astronomy Research Center, Seoul National University, 1 Gwanak-ro, Gwanak-gu, Seoul 08826, Korea}

\author[0000-0002-4391-2275]{Juhan Kim}
\affil{Center for Advanced Computation, Korea Institute for Advanced Study,
85 Hoegiro, Dongdaemun-gu, Seoul 02455, Korea}

\author[0000-0002-5751-3697]{M. James Jee}
\affil{Department of Astronomy, Yonsei University,
50 Yonsei-ro, Seoul 03722, Korea}
\affil{Department of Physics, University of California,
Davis, One Shields Avenue, Davis, CA 95616, USA}

\author[0000-0003-4032-8572]{Hyowon Kim}
\affil{Korea Astronomy and Space Science Institute (KASI),
Daedeokdae-ro, Daejeon 34055, Korea}
\affil{University of Science and Technology (UST),
Gajeong-ro, Daejeon 34113, Korea}

\author[0000-0001-5303-6830]{Rory Smith}
\affil{Korea Astronomy and Space Science Institute (KASI),
Daedeokdae-ro, Daejeon 34055, Korea}



\begin{abstract}
In a galaxy cluster, the relative spatial distributions of dark matter, member galaxies, gas, and intracluster light (ICL) may connote their mutual interactions over the cluster evolution. However, it is a challenging problem to provide a quantitative measure for the shape matching between two multi-dimensional scalar distributions. We present a novel methodology, named the {\em Weighted Overlap Coefficient (WOC)}, to quantify the similarity of 2-dimensional spatial distributions. We compare the WOC with a standard method known as the Modified Hausdorff Distance (MHD). We find that our method is robust, and 
performs well even with the existence of multiple sub-structures. We apply our methodology to search for a  visible component whose spatial distribution resembled with that of dark matter. If such a component could be found to trace the dark matter distribution with high fidelity for more relaxed galaxy clusters, then the similarity of the distributions could also be used as a dynamical stage estimator of the cluster. We apply the method to six galaxy clusters at different dynamical stages simulated within the GRT simulation, which is an N-body simulation using the galaxy replacement technique. Among the various components (stellar particles, galaxies, ICL), the ICL+ brightest cluster galaxy (BCG) component most faithfully trace the dark matter distribution. Among the sample galaxy clusters, the relaxed clusters show stronger similarity in the spatial distribution of the dark matter and ICL+BCG than the dynamically young clusters. While the MHD results show weaker trend with the dynamical stages.

\end{abstract}



\section{Introduction} \label{sec:intro}

One of the most promising routes to understanding how a cluster assembles is to observe cluster quantities which are sensitive to the evolutionary state of the cluster. Recent deep observations of nearby clusters show distinct diffuse \textit{intracluster light (ICL)}, which is light from stars not gravitationally bound to any individual cluster galaxy \citep{2002ApJ...575..779F, 2004ApJ...617..879L, 2005ApJ...618..195G, 2005MNRAS.358..949Z, 2005ApJ...631L..41M, 2017ApJ...834...16M, 2014MNRAS.437.3787C,2019ApJ...871...24C, 2018MNRAS.474.3009D, 2018ApJ...862...95K, 2019A&A...622A.183J, 2020ApJS..247...43K, 2021MNRAS.502.2419F, 2021MNRAS.508.2634Y}.

With regard to galaxy cluster evolution, one may raise a question: how does the spatial distribution of dark matter vary with redshift, mass, and dynamical stage of the galaxy cluster? The gravitational weak lensing is widely applied to investigating the dark-matter distribution in clusters.
Although it is powerful, the weak lensing method also has several drawbacks such as the difficulty of identifying background galaxies and accurately measuring their shapes in the presence of the telescope point spread function. Therefore, it would be worthwhile to search for other quantities that can recover the dark matter distribution, which could provide complementary information to enhance our understanding.

Unlike the hot gas component, the ICL is collisionless, and at the same time gravitationally bound to the global potential of the galaxy cluster (not to individual galaxies). In the current $\Lambda$CDM paradigm, dark matter halos are built up by hierarchical assembly of smaller halos, and the ICL grows through the assembly history of galaxy clusters \citep{2005ApJ...631L..41M, 2016IAUS..317...27M}.
Accordingly, ICL is expected to map the spatial distribution of dark matter, and several previous studies have shown it does \citep{2019MNRAS.482.2838M, 2020MNRAS.494.1859A}.
 Moreover, the ICL could be used as an effective tool to measure the evolutionary stage of galaxy clusters by assuming that the more evolved clusters would contain a richer ICL component, and that its spatial distribution would be well aligned with the distribution of dark matter.
 Making a rigorous statistical comparison of the spatial distributions of ICL and dark matter would be a crucial step in the study of ICL as a \textit{luminous tracer for dark matter}, and as an \textit{evolutionary stage estimator for galaxy clusters}.

The one-dimensional radial profiles of ICL and dark matter were compared in \citet{2021MNRAS.501.1300S}. However, the azimuthally-averaged radial profile did not fully represent the actual spatial distribution in 2- or 3-dimensions and thus consequently lost some valuable information.
Previous studies of the two-dimensional spatial distribution of ICL and dark matter \citet{2019MNRAS.482.2838M, 2020MNRAS.494.1859A} used the Modified Hausdorff Distance (MHD) method \citep{576361}, which quantifies the difference between two nearby contours using the distances between  points on the contours. The cross-correlation technique was adopted by \citet{2014ApJ...797..106H} to compare the dark matter maps from weak lensing analysis and galaxy cluster redshift surveys.
Outside  of the astrophysics field, the environmental niche overlap \citep{schoener1968anolis, warren2008environmental} is used in ecology, which shows the co-existence (overlap of resource use) of different species.
The Earth mover's distance \citep{Kantorovich}, also referred to as the Wasserstein metric, has gained recent popularity as a loss measure in certain machine learning applications involving image generation or classification.

In this particular statistical analysis, we compare two spatial distributions while the specific relation between their signal strengths is either not known {\em a priori} or is not being sought. In this work, we suggest that under such a specific requirement the aforementioned methods could suffer from biases and may not be the most appropriate choice. We thus introduce a new method for comparing the morphology of fields, and compare it with one of the previous methods, the MHD method. 

Although our method may be generalizable to a range of problems here we will restrict our analysis to the quantification and comparison of simulated massive galaxy clusters.
Furthermore we will investigate the ICL versus dark matter spatial distribution comparison and galaxy cluster dynamics. A similar comparison from our group, but focusing more on the large-scale spatial distribution in and around clusters based on a parametric method is presented in Shin et al. (in preparation). 

In \S2 we define the simulated data we will use.
In \S3 we introduce our new methodology and discuss its practical application. 
Our analysis on our method is presented in \S4, where we directly compare our new method with the MHD method. We discuss about the results and possible applications in \S5.
We summarize in \S6.

\section{Simulation Data}
\begin{figure*}
\begin{center}
\includegraphics[width=0.9\columnwidth,trim=3cm 3.9cm 1cm 4.2cm,clip]{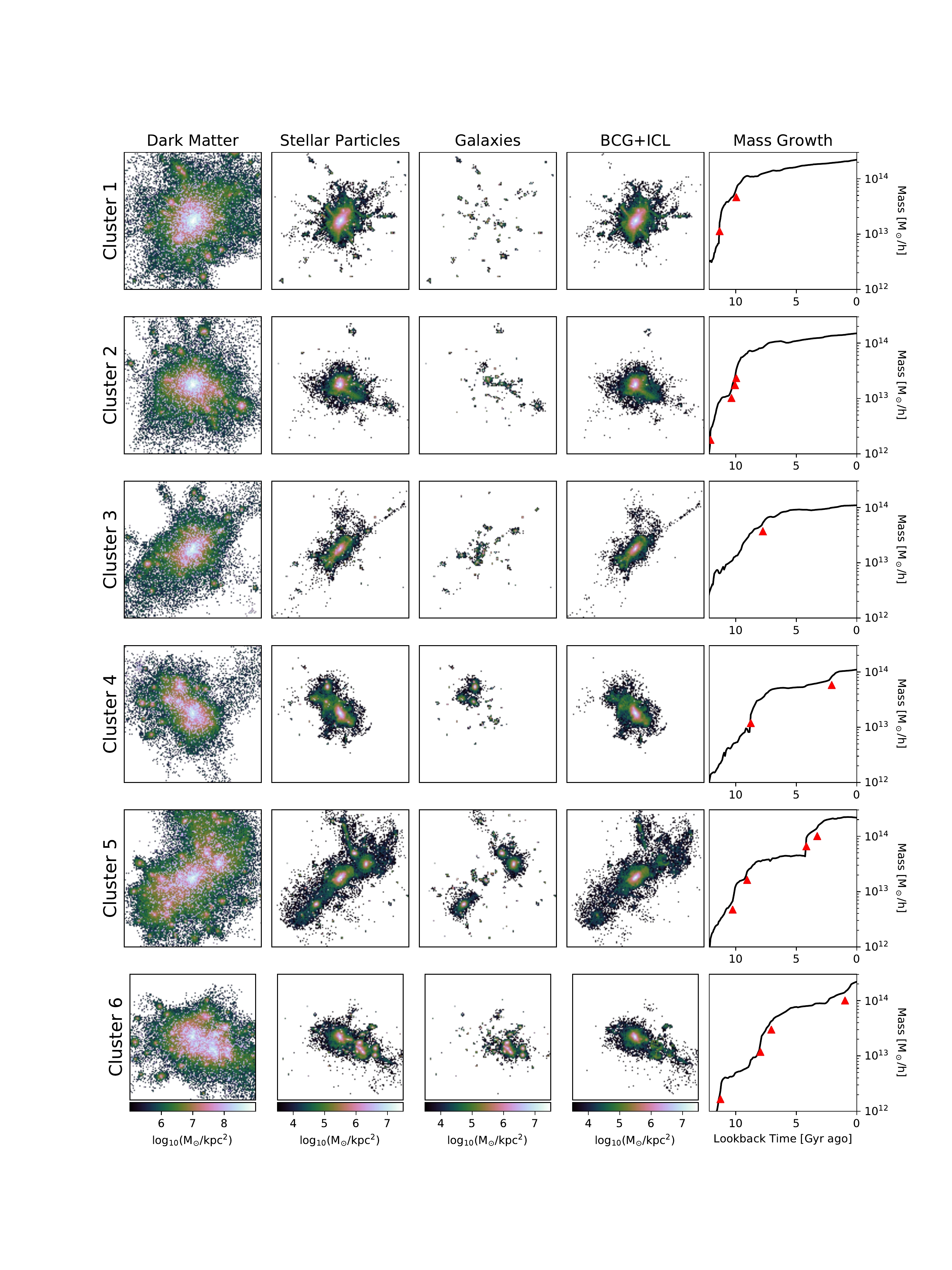}
\caption{Galaxy clusters simulated in the GRT simulations. Projected images in 2.5 $\times$ 2.5 $h^{-2}$Mpc$^2$ boxes for the dark matter (first column), all stellar particles (second column), member galaxies excluding BCG (third column) and BCG+ICL (fourth column) are shown. Throughout this work, the spatial distribution comparison between dark matter and galaxies is conducted using all member galaxies including the BCG, however we show satellite galaxies without the BCG as an aid the visualize the relative distributions of the galaxies and BCG+ICL. The mass growth histories of each cluster are plotted in the fifth column. The major merging events (mass ratio smaller than 1:5) are marked as red triangles.}
\label{fig:Ncluster}
\end{center}
\end{figure*}

We applied our method to six galaxy clusters lying at different dynamical stages selected from Galaxy Replacement Technique \citep[GRT,][]{2022ApJ...925..103C} simulation, which is run with the Gadget3 code \citep{2005MNRAS.364.1105S} for the study of cosmological gravitational evolution of galaxy clusters.
First, we performed the N-cluster Run, a low-resolution dark matter only simulation with $512^3$ particles of a uniform mass ($10^9~ h^{-1}{\mathrm M_\odot}$) in a cubic box of a side length, $120 ~ h^{-1}$Mpc.  Then, we resimulate the N-cluster Run but in a higher-resolution mode (10~$h^{-1}$pc), placing a stellar disk with high-resolution star particles according to the exponential disk galaxy model \citep{2009MNRAS.396..121D}. 
We replace only the dark matter halos that fall into the galaxy cluster and contribute to its growth. When a dark matter halo in the low-resolution simulation reaches its maximum mass as a central halo, and if the mass is greater than $10^{11}~ h^{-1}{\mathrm M_\odot}$, the dark matter halo is replaced by a high-resolution dark matter halo and high-resolution stellar disk. Since the replacement happened before the halo enters the virial radius of the host cluster, we assume it does not suffer from significant tidal interactions and does not contribute to the BCG+ICL. As these high-resolution halos fall into the clusters and are tidally stripped, the BCG and ICL form naturally.
For the GRT simulation, we use
the post-Planck cosmological model of $\Omega_m=0.3$, $\Omega_\Lambda=0.7$,
$\Omega_b=0.047$, and $h=0.684$.
Please refer to \citet{2022ApJ...925..103C} for a detailed description of the resimulation.
The GRT allows us to trace the tidal interaction histories of stellar particles in galaxies and galaxy clusters.
The high resolution of GRT simulations (low mass of star particles of 5.4$\times$10$^4~ h^{-1}{\mathrm M_\odot}$) allows us to resolve the low surface brightness features down to $\mu_V <$ 32 mag arcsec$^{-2}$. This surface brightness limit is approximately three mag arcsec$^{-2}$ lower than in Illustris TNG100 \citep{2018MNRAS.475..624N}.

The brightest cluster galaxy (BCG), other member galaxies, and ICL are defined among the stellar particles using the six-dimensional phase-space friends-of-friends halo finding algorithm, ROCKSTAR \citep{2013ApJ...762..109B}. We define the stellar particles which belong to the BCG and other member galaxies. The ICL is then defined as stellar particles which belong to the galaxy cluster but not to any member galaxies. We show the projected maps of each component in Figure \ref{fig:Ncluster}. From the first to the fourth column, maps of the dark matter and each component are shown. In the fifth column, the mass growth histories are shown, with the indication of the major merging events.
Considering the formation time ($z_{m/2}$; redshift at which 50\% of the cluster mass is obtained) and the redshift when the last major merger occurred ($z_{LMM}$), three dynamically young clusters (cluster 4, 5, and 6) and three dynamically relaxed fossil clusters (cluster 1, 2, and 3) were defined. Clusters 4, 5, and 6 have relatively recent major merging events. Specifically, cluster 6 seems to currently be undergoing major merging. Meanwhile, clusters 1, 2, and 3 had their last major merger events at much earlier times. Their formation times ($z_{m/2}$, see Table \ref{tab:table1}) were also relatively early.
We projected the simulated galaxy clusters in three different viewing angles (i.e., the $x-y$, $x-z$, and $y-z$ planes) to avoid biases arising from any one particular projection, augmenting our data sample. The projected images were cropped to a scale of 2.5 $\times$ 2.5 $h^{-2}$Mpc$^2$, which were then pixelized to 1024 $\times$ 1024 pixels. Thus, the pixel scale for the images is 2.44 $h^{-1}$kpc/pixel.

\begin{table}[t]
	\centering
	\begin{tabular}{ccccc}
		\hline Name & Mass & $z_{m/2}$ & $z_{LMM}$ & BCG+ICL \\
		&(${\mathrm M_\odot/h}$)&&&fraction (\%)\\
		 
		\hline 
		 cluster 1 &  $2\times 10^{14}$ & 1.386 & 1.715&60\\  
		 cluster 2 &  $1.5\times 10^{14}$ &1.106 & 1.766&65\\  
		 cluster 3 &  $10^{14}$ & 0.955 & 0.945&38\\  
		 cluster 4 &  $10^{14}$ & 0.384 & 0.154&49\\
		 cluster 5 &  $2\times 10^{14}$ & 0.353 & 0.276&33\\
		 cluster 6 & $2\times 10^{14}$ & 0.174 & 0.063 &25\\
		\hline 
	\end{tabular}
	\caption{ Cluster properties in GRT simulations. $z_{m/2}$: redshift at 50\% of the cluster mass($z=0$) obtained. $z_{LMM}$: redshift when the last major merger happened.}
	\label{tab:table1} 
\end{table}

\section{Methods} \label{sec:method}
\subsection{WOC method} \label{subsec:WOC}
\begin{figure}
\begin{center}
\includegraphics[width=0.7\columnwidth,trim={1.5cm 0.5cm 1.5cm 1.5cm},clip]{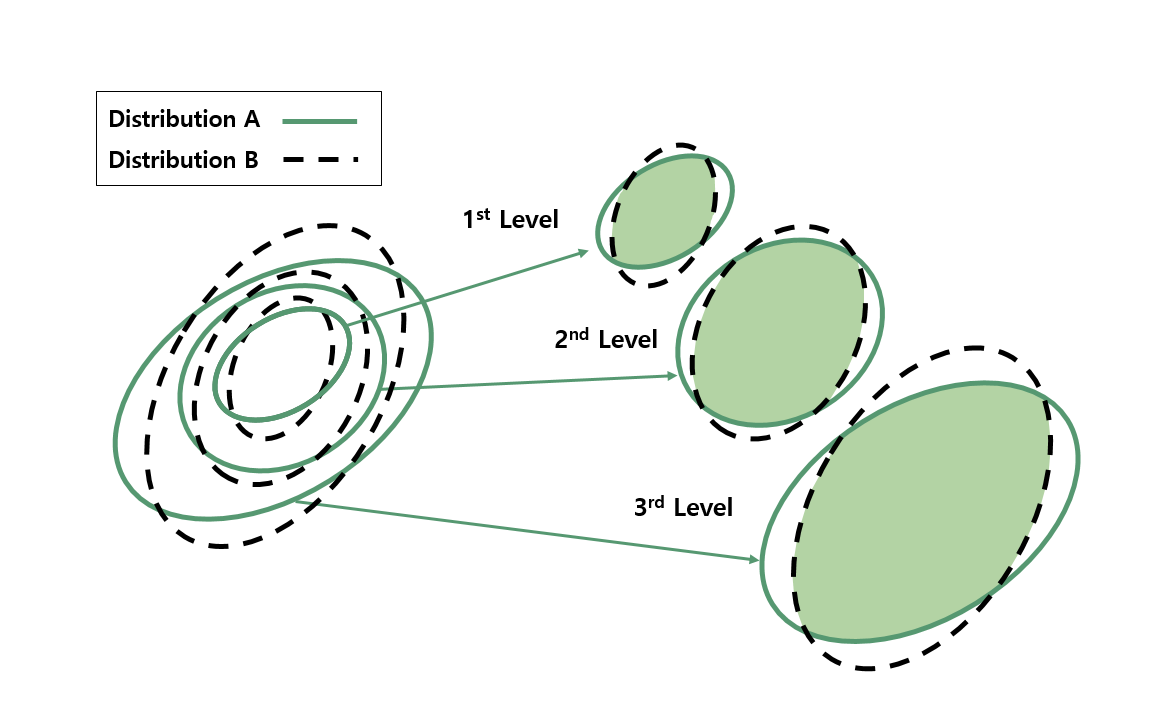}
\caption{WOC method idea. We measure the overlap area of contours between two distributions at various density threshold levels. We calculate the fraction of the overlapping area and give weight using the signal strength to quantify their overall similarity as a normalized number between 0 and 1. See more details in Section \ref{subsec:WOC}.
\label{fig:woc_idea}}
\end{center}
\end{figure}

\begin{figure*}
\begin{center}
\includegraphics[width=0.8\columnwidth,trim={2cm 7.5cm 2.2cm 0},clip]{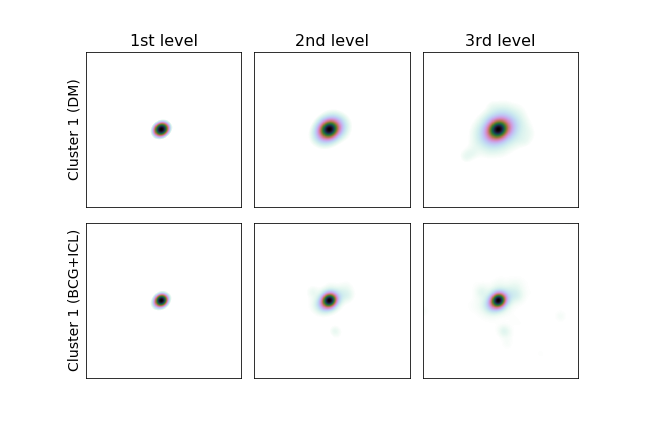}
\includegraphics[width=0.8\columnwidth,trim={1.8cm 2cm 2cm 2.6cm},clip]{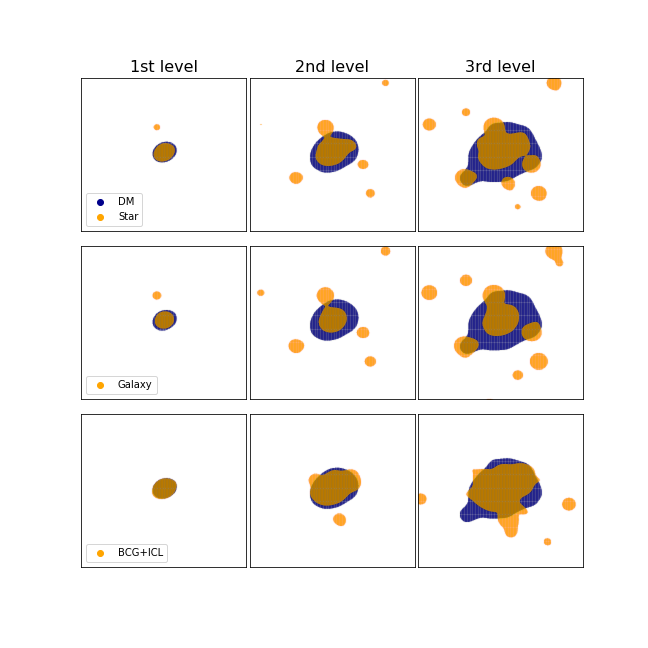}
\caption{WOC method measuring process for  cluster 1, simulated in GRT simulations. All panels have a box size of 1.5 $\times$ 1.5 $h^{-2}$Mpc$^2$. The dark matter distribution in cluster 1 (first row) was compared with all stellar particles (second row; cluster galaxies including BCG+ICL), member galaxies (third row; cluster galaxies including BCG), and BCG+ICL (fourth row). For three lower panels, the distributions of dark matter and the other compared components are marked with blue and orange colors, respectively. We measured the overlap area for three levels, in the order from highest to lowest, as the first (left), second (middle), and third (right) levels, and gave them weight to calculate the WOC.
\label{fig:process}}
\end{center}
\end{figure*}


\begin{figure*}
\begin{center}
\includegraphics[width=0.65\columnwidth,trim={2.2cm 1.6cm 2.2cm 1.2cm},clip]{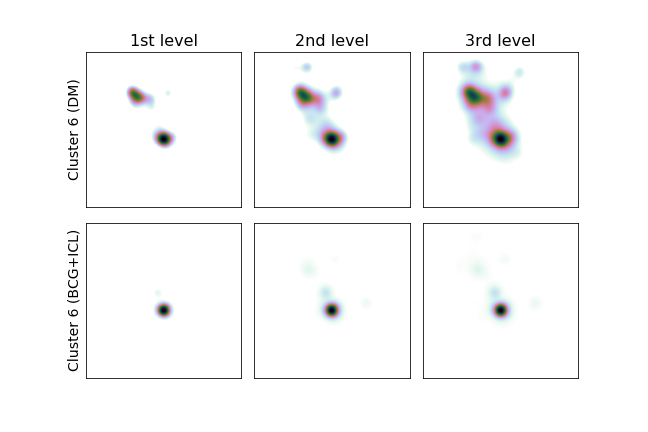}
\includegraphics[width=0.65\columnwidth,trim={2cm 0.8cm 2cm 0.8cm},clip]{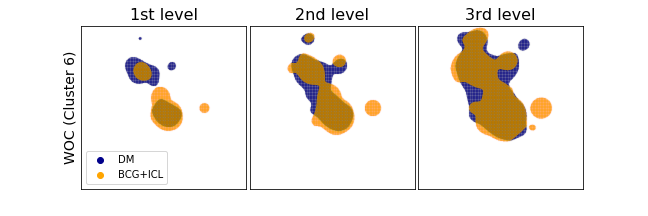}
\includegraphics[width=0.65\columnwidth,trim={2cm 0.8cm 2cm 0.8cm},clip]{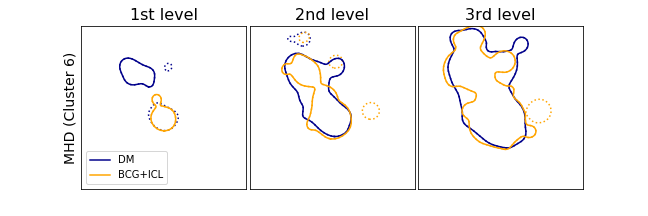}
\includegraphics[width=0.65\columnwidth,trim={2cm 0.8cm 2cm 0.8cm},clip]{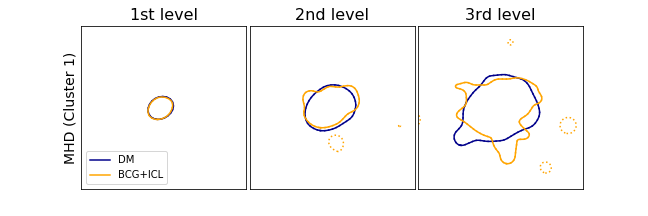}
\caption{The intermediary steps used for measurement with the  WOC (third row) and MHD (fourth and fifth row) methods, for cluster 6 (third and fourth row) and cluster 1 (fifth row). The raw data images of dark matter and BCG+ICL of cluster 6 are plotted in the first and second row, respectively. All of the panels have a box size of 1.5 $\times$ 1.5 $h^{-2}$Mpc$^2$. For both methods, the distributions of dark matter and BCG+ICL are marked with blue and orange colors, respectively. The three levels of dark matter and BCG+ICL distributions are plotted in the left (first level), middle (second level), and right (third level) panels. In the MHD method, when there were several contours for one level (marked by the dotted line), the largest contour (solid line) was selected to compare with the contour of the other component.
\label{fig:process2}}
\end{center}
\end{figure*}
We introduce a novel method to quantify the similarity of two 2-dimensional spatial distributions. The similarity is quantified by the \textit{Weighted Overlap Coefficient (WOC)} whose functional form is given in equation \ref{eq:woc}. In this, we calculate the fraction of overlapping area between two distributions at various density threshold levels (see Figure \ref{fig:woc_idea}). Using a weighted sum of those fractions, we normalize the similarity measure between 0 and 1.


The calculation proceeds as follows:
\begin{enumerate}
    \item Prepare two-dimensional maps of two components smoothed on the same angular or spatial scale.
    \item Determine the signal  center in the reference map (for example, the peak position of the dark matter density map) and measure the one-dimensional radial profile. Find the signal levels at $n$ equally-spacing radii (for example, if $n=3$, $r=100$, 200, and 300 kpc).
    \item Find the enclosed area at each level of the isolevel contour (i.e. pixels with signal values larger than the input signal levels). When there are multiple disconnected contours in the map or multiple isolated areas, simply add up all the areas.
    \item In the comparison map (for example ICL), find contour levels to have the same enclosed area size as those found in the reference map. We do this by starting from the peak signal and lowering the pixel value threshold in small steps until the area constraint is satisfied. This step is needed to quantify the overlapping area as a percentage.
    \item Measure the amount of overlapping areas between the reference and comparison maps at each signal level and calculate the overlapping fraction (see Figure \ref{fig:process}).
    \item Parametrize the similarity of the two component maps with $n$ overlapping fractions multiplied by weightings as (1) the inverse of area fraction of each level to the lowest level ($w_{A}$), (2) the signal level  at the contour in the reference component ($w_{\rho_1}$), and (3) the signal level at the contour in the comparison component ($w_{\rho_2}$). See equation \ref{eq:woc} for the mathematical form. 
\end{enumerate}

Finally the similarity of the two spatial distributions is  quantified as the WOC, which is a single number between 0 and 1.
For two maps $I_1$ and $I_2$, the WOC is defined as
\begin{equation}
WOC(I_1,I_2)=\frac{ \sum\limits_{i=1}^{n} a_i \left(w_{A,i}+w_{\rho_1,i}+w_{\rho_2,i}\right)}{\sum\limits_{i=1}^{n} \left(w_{A,i}+w_{\rho_1,i}+w_{\rho_2,i}\right)}
\label{eq:woc}
\end{equation}
with
\begin{equation}
\sum_{i=1}^{n} w_{A,i}=\sum_{i=1}^{n} w_{\rho_1,i}=\sum_{i=1}^{n} w_{\rho_2,i}=1 ,
\label{eq:weight}
\end{equation}
where $a_i$ is the overlap area (\%) for the $i^{th}$ level, $w_{A,i}$ is the area ratio of the $i^{th}$ contour compared to the largest ($n^{th}$) contour, $w_{\rho_1,i}$ is the density level  at the $i^{th}$ contour on the Map 1 and $w_{\rho_1,i}$ is the density level at the $i^{th}$ contour on the Map 2.

Throughout this weighting system, we take account of both the area of each contour, which are used to measure the overlapping region, and the strength of the signal of the selected contours of each map (i.e., dark matter and ICL). Among the contours on various levels, the contour for the stronger signal level would occupy smaller area than that of the weaker signal. Moreover, if the distribution is rough, the contour of the stronger signal would be smaller than the smooth distribution situation at the same level. If the sharp peaks of two distributions match well with each other, even though the chance of it would be smaller due to the small area, then we could regard those cases as {\em the location of two distributions coincide}. Therefore, using the weight $w_{A}$ (area ratio), we give more weight to smaller contours. The contour areas of two distributions for the same level to compare are set to be same, thus $w_{A}$ would represent the signal strength of both distributions. The signal strengths of the individual distributions are taken into account more directly/ separately, via their density value $w_{\rho_1}$ and $w_{\rho_2}$. If the distribution maps are density fields, $w_{\rho_1}$ and $w_{\rho_2}$ would be simply the pixel value (before normalization in equation \ref{eq:weight}) at the contour to compare on the Map 1 and Map 2, respectively.

Note that, we are not interested in the relative signal strengths of each component or the exact shape of their profiles, rather we focus on the spatial correspondence of the two components. One of the main advantages of this method is that it performs well when comparing two distributions when either or both contain disconnected regions.
Furthermore, in the WOC method we do not need to compute the individual contours, and thus it suffers less bias even when working with masked maps.

\subsection{MHD method\label{subsec:MHD}}
Previous studies \citep{2019MNRAS.482.2838M, 2020MNRAS.494.1859A} of the two-dimensional spatial distribution of ICL and dark matter have used the MHD method. 

The MHD defined by \cite{576361}:
\begin{equation}
d_{MH}(X,Y)=\max(d(X,Y),d(Y,X)),
\end{equation}
where
\begin{equation}
d(X,Y)=\frac{1}{N_X} \sum\limits_{x\in X} \min_{y\in Y}\parallel x-y\parallel.
\end{equation}
The two sets of points, $X=\{x_1, x_2, ..., x_{N_x}\}$ and $Y=\{y_1, y_2, ..., y_{N_y}\}$, define two contours, and $\parallel\cdot\parallel$ is the Euclidean norm.
To compare the MHD between different clusters, we adopt the relative MHD defined by \cite{2020MNRAS.494.1859A}:
\begin{equation}
\zeta=\frac{d_{MH}(X_r,Y_r)}{r},
\label{eq:mhd}
\end{equation}
for the distance $r$, which also in turn defines the contours, $X_r,Y_r$, via the angle averaged one-dimensional radial profile.
Situations can arise where a level defined by $r$ results in more than one contour, i.e., multiple disconnected regions (see Figure \ref{fig:process2}). In this case we follow the suggestion of \citet{2020MNRAS.494.1859A} and choose only the largest contour when evaluating equation \ref{eq:mhd}.
Thus, this method may not be robust in cases where the spatial distributions exhibit multiple local sub-structures (disconnected from the main body), which are expected in dynamically active clusters, such as Bullet or Coma cluster. 
For example, the previous study using the MHD method \citep{2020MNRAS.494.1859A} excluded samples from their analysis, if there are more than one central galaxy in the cluster.

\section{Analysis} \label{sec:analysis}
\subsection{Modified WOC method and Characteristics of different methods}\label{sec:Character}

\begin{figure*}[p]
\begin{center}
\includegraphics[width=1.0\columnwidth,trim={2cm 1cm 3cm 0cm},clip]{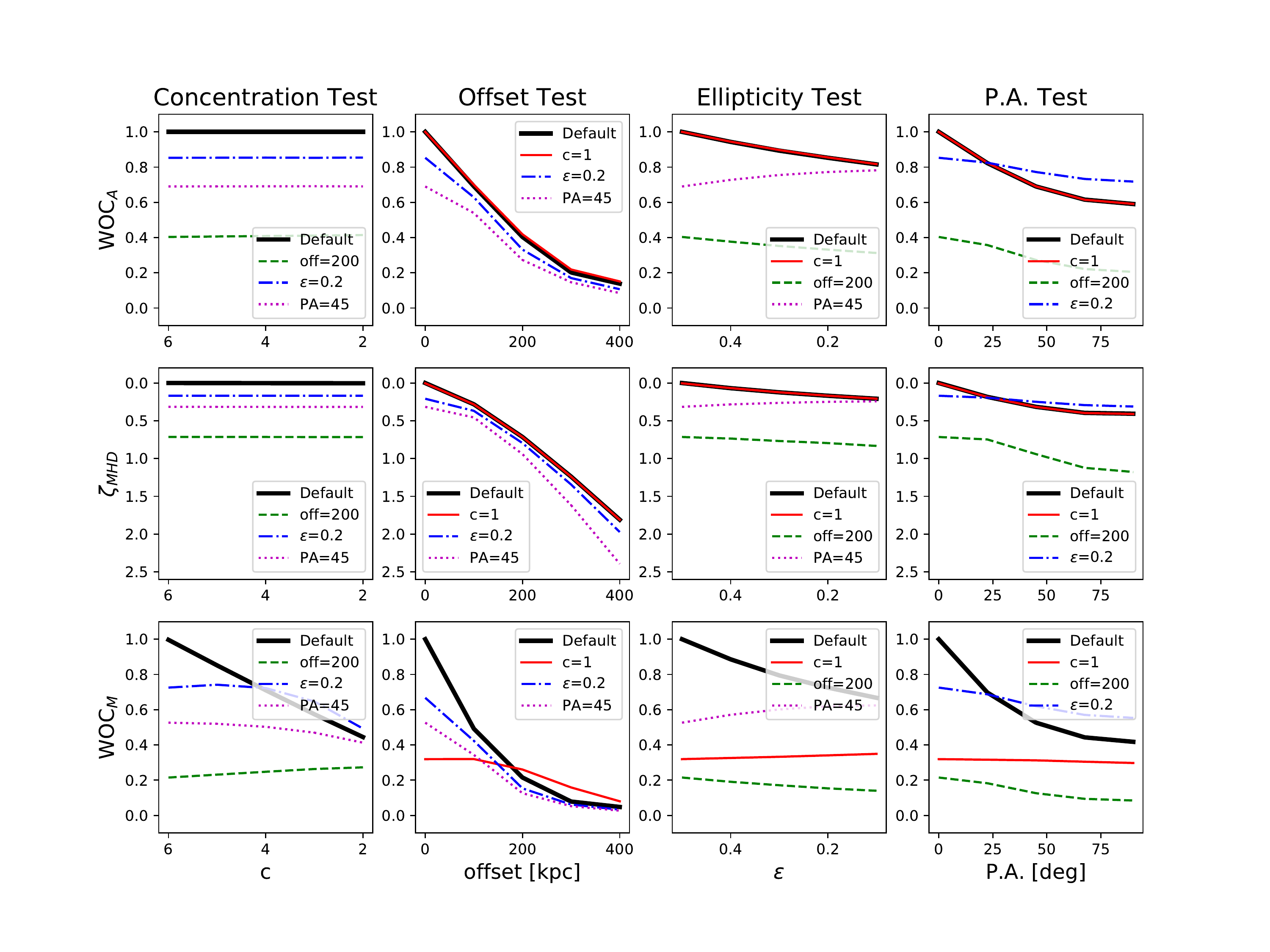}
\caption{Characteristics of the original WOC (upper panel), MHD (middle panel) and  WOC$_M$ (lower panel) methods for varying parameters of compared NFW mock image. Fixing the four parameters for the reference distribution (Map 1 in Table \ref{tab:table_NFW}), varying the concentration rate (first column), offset (second column), ellipticity (third column), and PA (fourth column) of the Map 2 results  are shown (thick black line). For each plot, taking a different concentration (red line), offset (green line), ellipticity (blue line), and PA (magenta line), the values of the Map 2 cases are overplotted. The parameter sets of the mock images are summarized in Table \ref{tab:table_NFW}.
\label{fig:param}}
\end{center}
\end{figure*}
\begin{table*}[p]
	\centering
	\begin{tabular}{c|c|cccc}
		\hline 
		Parameter&& Concentration Test & Offset Test & Ellipticity Test & P.A. Test\\ 
		 & Map 1 & Map 2 &  Map 2 & Map 2 & Map 2 \\
		
		\hline Concentration & 6 & 6, 5, 4, 3, 2 & 6/1 & 6/1 & 6/1 \\
		 Offset [kpc]&  0 &0/200 & 0, 100, 200, 300, 400 & 0/200 & 0/200\\  
		 Ellipticity & 0.5 & 0.5/0.2 & 0.5/0.2 & 0.5, 0.4, 0.3, 0.2, 0.1 & 0.5/0.2\\  
		 P.A. [deg]& 0 & 0/45 & 0/45 & 0/45 & 0, 22.5, 45, 67.5, 90 \\  
		\hline 
	\end{tabular}
	\caption{Parameters used for WOC$_A$,  MHD and  WOC$_M$ test using mock NFW halo images (See Section \ref{sec:Character} and Figure \ref{fig:param}). For each parameter test, Map 1 with default parameter values and Map 2 with varying designated parameters were compared.}
	\label{tab:table_NFW} 
\end{table*}
What does it mean for one distribution to \textit{trace} another distribution? In some cases we may want to quantify both the spatial co-existence and  the corresponding signal strength together. The WOC method, however,  focuses solely on  the spatial mismatch between two distributions, and is not sensitive to the relative strength difference within distributions, such as the shape of the radial profile. This is because we set the contours using the {\em reference} distribution, and find the corresponding contours with the same areas from the {\em comparison} distribution. 

We developed another WOC system (WOC$_M$), where the contours from the comparison distribution are not found using the same area, but rather by the same mass fraction. For example, under the WOC$_M$ method, if the contour of the first level for the dark matter distribution encloses 10\% of the total mass of dark matter, it will be compared with a contour which encloses 10\% of the total mass of the comparison component. In this case, the overlap area $a_i$ (\%) in equation \ref{eq:woc} will be calculated as the overlapped region area divided by the total area occupied by the contours. Here, the signal strength is already taken into account by setting the contours, thus, among the three weight factors in equation \ref{eq:woc}, we have only one weight factor, which is the area ratio, $w_{A,i}$.

We generated mock images that followed NFW profiles, while varying four parameters; the concentration rate, offset, ellipticity, and position angle (PA). We then checked how the different spatial distribution comparison methods reacted to the various situations (see Figure \ref{fig:param} and Table \ref{tab:table_NFW}). Fixing the four parameters for the reference distribution (Map 1 in Table \ref{tab:table_NFW}), varying the concentration rate (first column), offset (second column), ellipticity (third column), and PA (fourth column) in the Map 2 results by the original WOC (upper row), WOC$_M$ (middle row), and MHD (lower row) methods are shown (thick black line). For each plot, taking a different concentration (red line), offset (green line), ellipticity (blue line), and PA (magenta line), the values in the Map 2 cases are overplotted.

The original WOC method (WOC$_A$) and the MHD method ($\zeta_{MHD}$) showed similar results, which were insensitive to the concentration rate, but were sensitive to the other three parameters changes. The results of the modified WOC method (WOC$_M$) dropped as the concentration rate in Map 2 differed from Map 1. The WOC$_M$ method is highly sensitive to the concentration rate (radial profile) change, so that it could not detect the offset, ellipticity, and PA difference if the comparing distribution had a different concentration rate (the red lines in the middle row in Figure \ref{fig:param}). Thus, the results of the modified WOC method could exhibit a degeneracy between the situations where cluster components have different profiles or where they are spatial mismatched, in the low WOC$_M$ case. However, the WOC$_M$ method has an advantage over WOC$_A$ method in that it compares contours with the same mass ratio (more physical meaning assigned to the pair of contours) and has a simplified weighting system. Depending on their scientific interest, ones could choose an appropriate method.

\subsection{Measurement of Overlap Area before binning and weighting}\label{sec:overlapTest}

\begin{figure}
\begin{center}
\includegraphics[width=0.55\columnwidth,trim={0 0 1.5cm 0},clip]{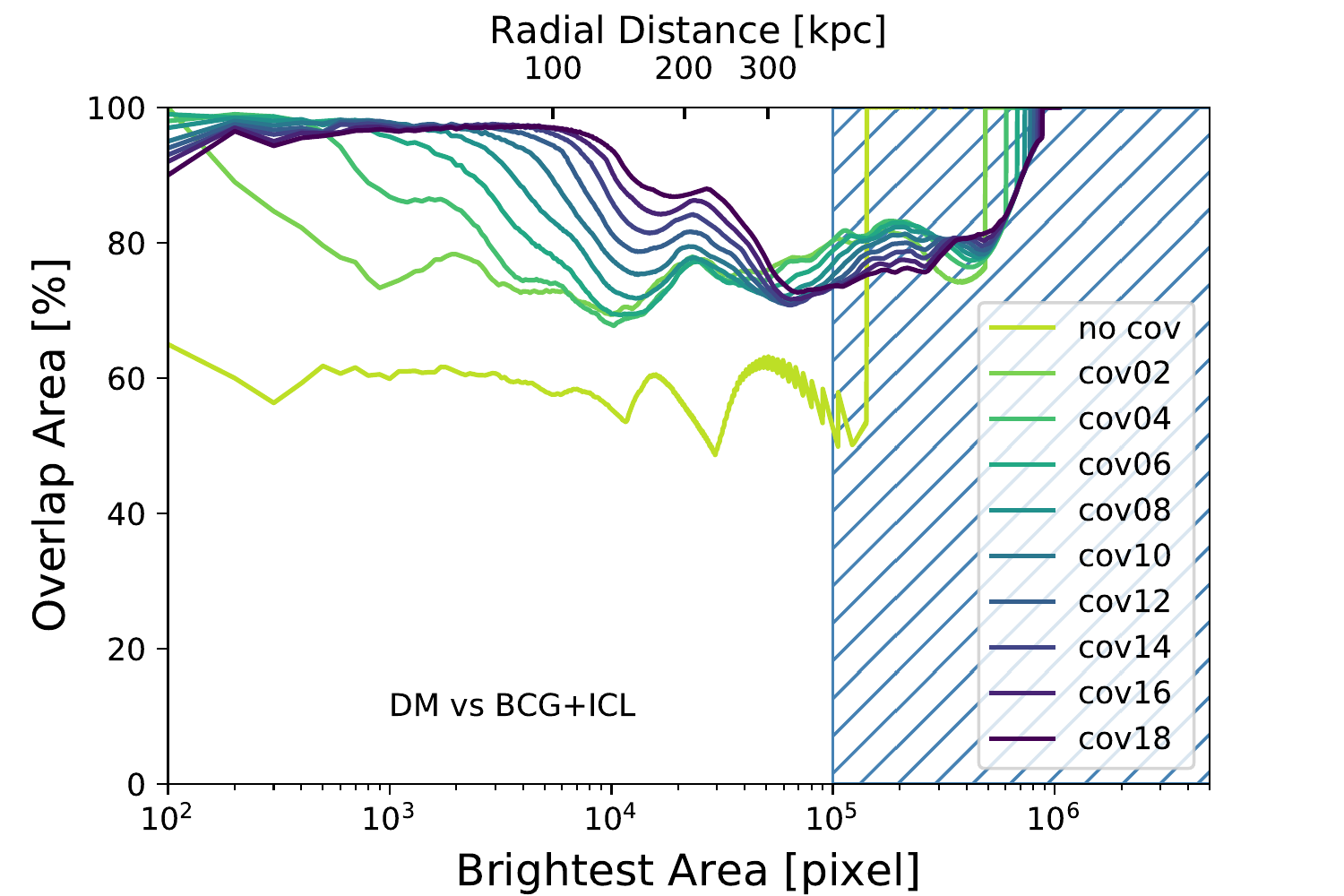}
\includegraphics[width=0.55\columnwidth,trim={0 0 1.5cm 1.1cm},clip]{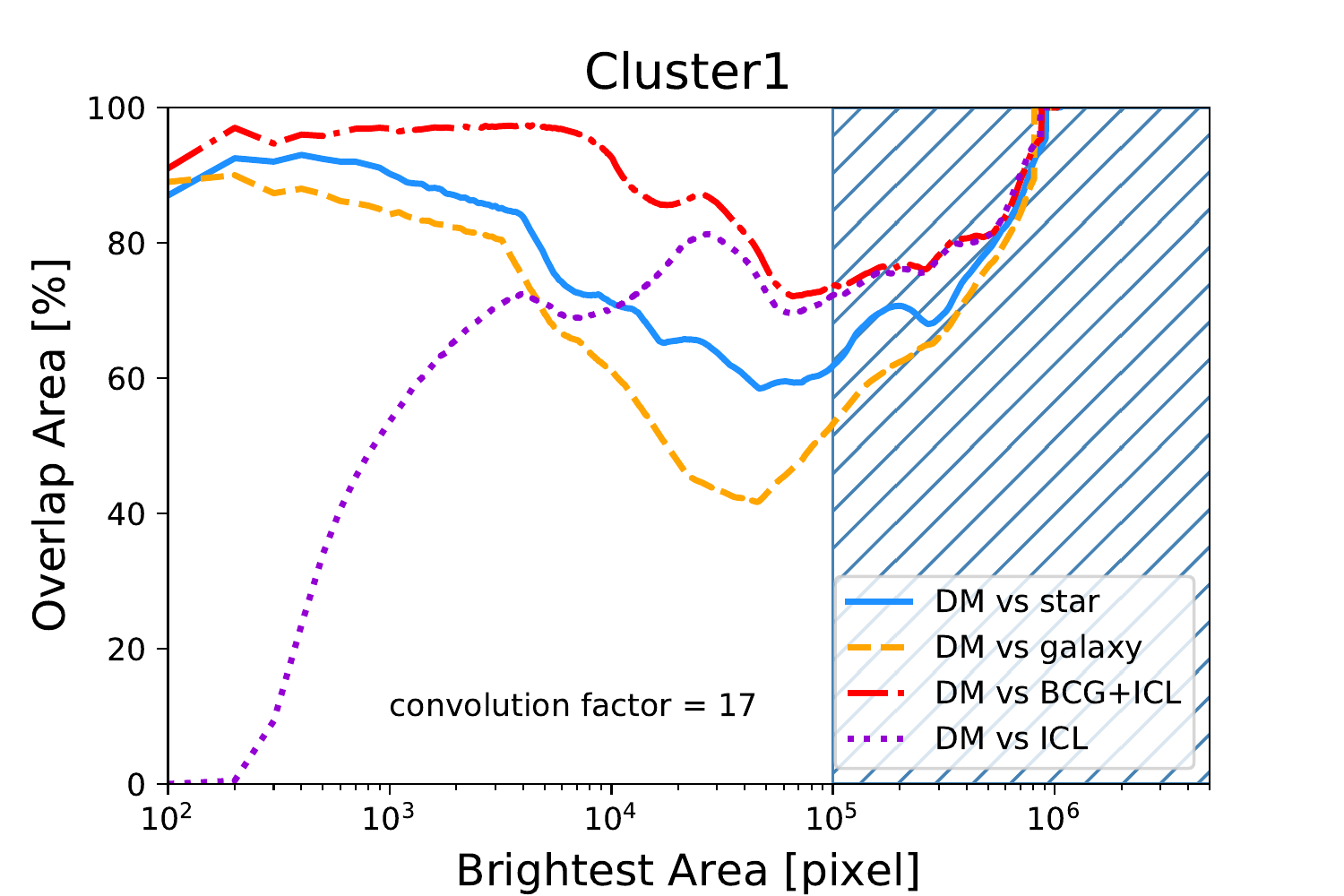}
\includegraphics[width=0.55\columnwidth,trim={0 0 1.5cm 1.1cm},clip]{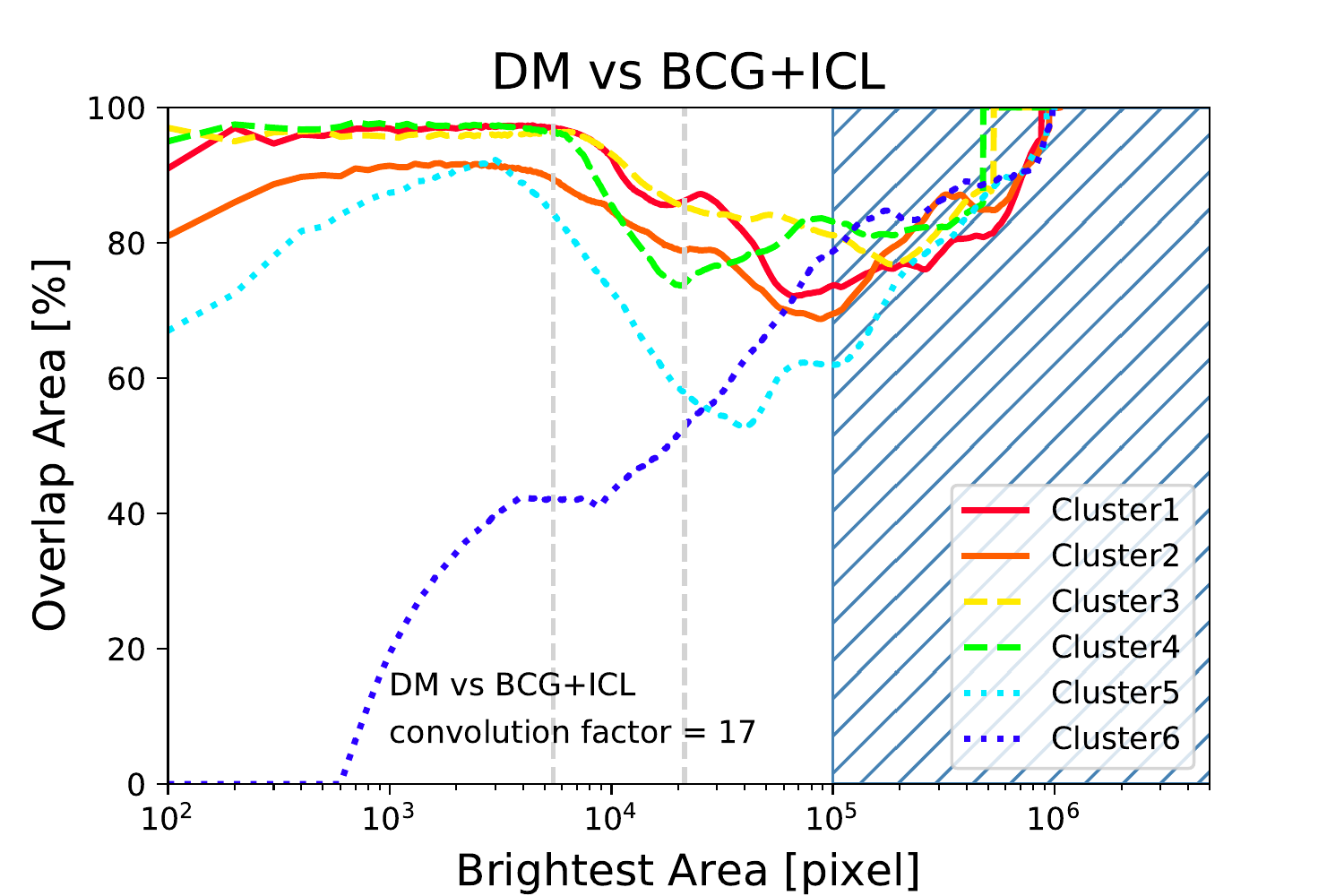}
\caption{Overlapping area fraction as a function of brightest area. Results for different convolution factors for dark matter vs. BCG+ICL of cluster 1 (upper panel), dark matter vs. different components of cluster 1 (middle panel), and dark matter vs. BCG+ICL of different galaxy clusters (lower panel) are plotted. For the x-axis (brightest pixel area), the corresponding physical scales of the radial profile of dark matter are marked together in upper panel.  Additionally, the pixel areas for 100 kpc and 200 kpc contour levels are marked as vertical gray dashed lines in the lower panel. The upturn at the largest area is an artifact of the image size and zero values in the images, which is marked as blue hatched region.}
\label{fig:cross}
\end{center}
\end{figure}
The WOC method measures by choosing a certain binning in the radial distance from the center of the image, measuring the overlap area percentage and giving weight. Without binning and weighting, we re-arranged the pixels of each distribution in the order of pixel value, and we then measured the overlap area while increasing the selected brightest area pixels. For example, we checked the location of the brightest 100 pixels in Map 1 and Map 2, and counted how many pixels resided in the same position. This measurement was repeated with increasing brightest area (see Figure \ref{fig:cross}). The maximum area was 1024 $\times$ 1024 pixels. The test was conducted for various image convolution factors (upper panel), various galaxy cluster components (middle panel), and various galaxy clusters with different dynamical stages (lower panel). 

 \textit{\bf Convolution factors}:  We vary the convolution scale in increments of 2 pixels from 0 upto 18 pixels. Depending on the convolution factor the behavior of the overlap area can change significantly, although at larger scales the curves become similar, as we might expect, since convolutions are local transformations.
 Because  Map 2 (in this case, the BCG+ICL of cluster 1) has pixels with a pixel value of zero, the overlap area for the non-convolved (the lime green line in the upper panel) and the lightly convolved images saturate at a certain point.
 
 \textit{\bf Cluster components}: For various components of an example galaxy cluster (cluster 1), we measured the overlap area with the dark matter distribution by increasing the brightest area (middle panel). The ICL only component (magenta dotted line) had zero overlap until the 200 brightest pixels. Apart from the ICL only component, all stellar particles (BCG + other member galaxies + ICL), all member galaxies, and the BCG+ICL components showed a high overlap area percentage until around 4000 brightest pixels. 
 As the brightest area grows (lowering the pixel value), new islands may emerge for each component and if they are  non-overlapping the overall percentage of overlap area will decrease, however this may recover again, resulting in the small dips in we see in Figure \ref{fig:cross}.
 Between the 5 $\times$ $10^3$ and 5 $\times$ $10^5$ interval, the BCG+ICL component (red dash-dotted line) showed a higher overlap than the stellar particles (blue solid line) or the member galaxies (yellow dashed line).
 Due to the fixed size of the images, and the fact that the faintest pixels all have a value of zero, the overlap area rise up again after the brightest area of $10^5$ pixels.
 
\textit{\bf Dynamical stages}: The overlap areas of the BCG+ICL and dark matter for the galaxy clusters with various dynamical stages are shown in the lower panel in Figure \ref{fig:cross}. Overall, the more relaxed clusters with earlier formation times (clusters 1, 2, and 3) had higher overlap than the dynamically younger clusters with later formation times (clusters 4, 5, and 6). For the case of cluster 6, the peaks of dark matter and BCG+ICL did not match, which caused a zero overlap until the 500 brightest pixels. 
Moreover, apart from cluster 6 (with a large offset between the dark matter peak and the BCG), the more relaxed clusters (clusters 1, 2, and 3) exhibit a flatter trend than the dynamically younger clusters (clusters 4 and 5) between the area interval corresponding to the 100 kpc and 200 kpc contours (see the vertical gray dashed lines in the lower panel of Figure \ref{fig:cross}). The ratios of the overlap area calculated for small and large contours (overlap area (\%) at 200 kpc contour/ overlap area (\%) at 100 kpc contour) are 0.89, 0.88, 0.88, 0.77, 0.68, and 1.25 for cluster 1, 2, 3, 4, 5, and 6, respectively. This may be expected from the \textit{fast collapse} and \textit{slow growth} scenario \citep{2015ApJ...810...36M, 2018ApJ...863...37F}, where both relaxed and unrelaxed system have well matched distribution between dark matter and BCG+ICL in the inner radius, while only relaxed systems have had enough time to  virialize in the outer regions. Thus, this ratio or the steepness of the slop in the lower panel of Figure \ref{fig:cross} could be used as an indicator of the dynamical stage. Even cluster 6 could be an example use case for this indicator, showing that the ratio of overlap area fractions for small and large contours greater than one could indicate that the cluster is undergoing extreme merging events.
\subsection{Robustness Test} \label{sec:robust}
\begin{figure*}
\begin{center}
\includegraphics[width=0.95\columnwidth,trim={0 0 1.2cm 0.8cm},clip]{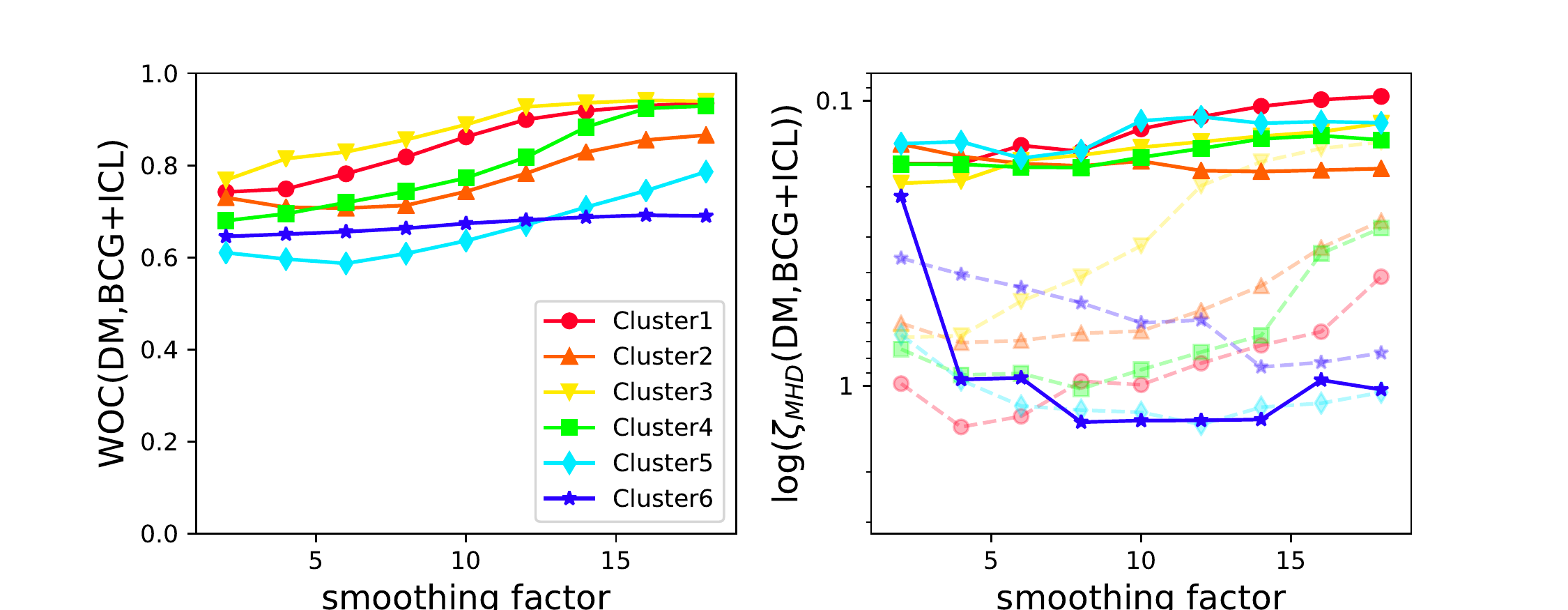}
\includegraphics[width=0.95\columnwidth,trim={0 0 1.2cm 0.5cm},clip]{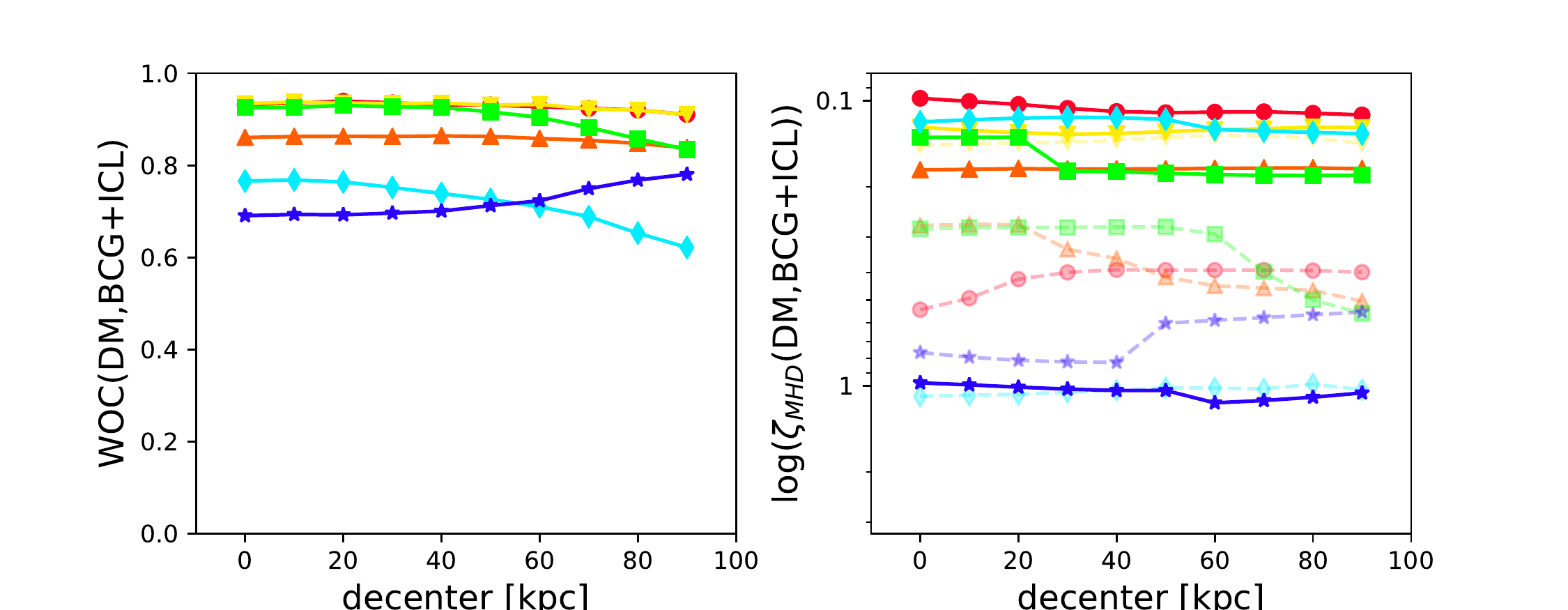}
\includegraphics[width=0.95\columnwidth,trim={0 0 1.2cm 0.5cm},clip]{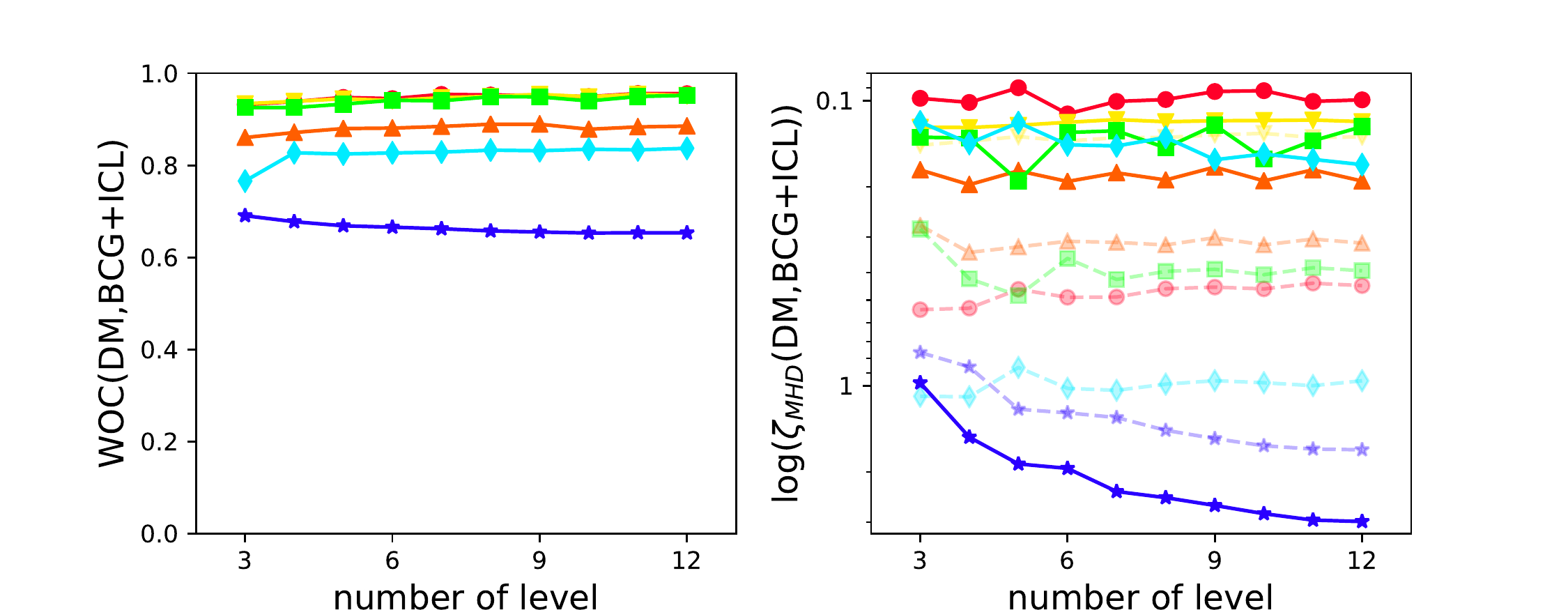} 
\caption{ The results of WOC(DM,BCG+ICL) (left) and $\zeta_{MHD}$(DM,BCG+ICL) (right) of different galaxy clusters for different smoothing factor (upper panel),  decenter (middle panel) and binning (lower panel). A smoothing factor of 17 is used for the decenter and binning tests. The labels are in the order of formation time ($z_{m/2}$), where cluster 1 formed early (dynamically relaxed) and cluster 6 formed late (dynamically young). Note that, for spatially matched distributions, the WOC would be close to 1, and the MHD would be close to 0. The range of the MHD result is indefinite, i.e., the $\zeta$-value is unbounded for the case of non-matching distributions. We thus take the logarithm of $\zeta_{MHD}$, to allow for a qualitative visual comparison between the tends of the MHD and WOC methods. The MHD method results using the largest contour are plotted as solid lines, while using all contours are plotted as dashed faint lines. 
\label{fig:Smooth}}
\end{center}
\end{figure*}
An ideal estimator should reflect the desired property that is being investigated (in this case, show a clear trend between WOC and the dynamical stage of galaxy clusters) and at the same time, be robust under data processing parameters such as smoothing scale, centroid, bin size, etc. In this section, we test the robustness of the WOC method while varying the convolution factor for the images, the decenter distance, and number of levels in the  measurement of the WOC using  GRT simulations galaxy clusters, and compare the results with those of the MHD method.

In the MHD method, the procedure choosing the largest contour for each level and excluding all sub-structures (see Section \ref{subsec:MHD}) could produce larger error, thus we compute both cases, i.e., using the largest contour and all contours.

Note that, a direct comparison between the results of WOC and MHD is difficult because the range of the MHD result is indefinite (see the y-axis of right panel in Figure \ref{fig:Smooth}), whereas the WOC result is normalized between 0 and 1 (see the y-axis of left panel in Figure \ref{fig:Smooth}). Thus, it is only appropriate to focus on the relative difference between samples within each method individually and look for systematic trends. We then may safely concluded the relative robustness (or biases) of each method under the different observational circumstances mentioned above. 

\textit{\bf Smoothing}: We convolved the images using the \texttt{Gaussian2DKernel astropy package}, where the full-width half-maximum size of the gaussian kernel is indicated as the smoothing factor.
As the smoothing scale applied to the input images increases, the resultant WOC also increases because the convolution fills the gaps between pixels and results in higher overlap fractions. And then, it becomes flat at around a smoothing factor of 16 $\sim$ 18 (see the upper left panel in Figure \ref{fig:Smooth}). 
 On the upper right panel in Figure \ref{fig:Smooth}, the same test was conducted with the MHD method using the largest contour for each level (solid line) and all contours (dashed faint line). To express the MHD result (which is a function of radius) as a single number, we calculated the relative MHD ($\zeta_{MHD}$) by taking the mean of $\zeta$ values for the galaxy cluster. To compare the trend of the MHD result with the WOC result, we took the logarithmic scale of $\zeta_{MHD}$. As the smoothing factor increased, the relative MHD value using the largest contour changed slightly, except for cluster 6. For cluster 6, which undergoes merging, the relative MHD value fluctuated severely (see the blue solid line in the upper right panel in Figure \ref{fig:Smooth}). This symptom was alleviated as we take account all contours (dashed faint line), however the overall results become less robust.

\textit{\bf Decentering}: The first step in calculating the WOC or the MHD is to determine the center of the galaxy cluster and compute a radial profile. Depending on the choice of the cluster center, the contours could be altered, which could cause differences in the results for the WOC or MHD. We tested the effect of decentering on the WOC/ MHD result (see middle panel of Figure \ref{fig:Smooth}). Both the WOC and MHD result seemed to be relatively robust compared to other parameter changes. Taking all contours for the MHD calculations, the difference between cluster 6 and other clusters becomes milder, however the overall scatter becomes larger. 

\textit{\bf Binning}: The choice of binning for measuring the overlap area at each level (for example $n=3$ and $r=100$, 200, 300 kpc) is somewhat arbitrary, and could affect the final result. Considering the observational studies of galaxy clusters, we set the outermost radius as 300 kpc, and divided it into $n$ equally spaced bins ranging from $n=3 \sim 12$ (see the lower panel in Figure \ref{fig:Smooth}). Both the WOC method and MHD method showed consistent results with the binning change, with the exception of the MHD result for cluster 6. As the number of levels for measuring the overlap area increased, the $\zeta_{MHD}$ of the dark matter and BCG+ICL component in cluster 6 rapidly increased (contours became mismatched). 

In general, both methods seem to be robust against our tests (except for cluster 6 case in the MHD method), although both methods appear more sensitive to the smoothing size rather than to decentering or the number of radial bins.
For all three robustness tests between the galaxy clusters, the relaxed clusters (cluster 1, 2, and 3) had an overall higher WOC than the dynamically younger clusters (cluster 4, 5, and 6). However, the MHD results did not show significant trend with the dynamical stage of sample clusters. Only  cluster 6 showed an undesirable fluctuation (trend) as the smoothing factor (binning) changed. Modifying the MHD method, taking account of all contours (substructure), alleviates the difference between cluster 6 and the other clusters, although this ends up producing less robust results. However, the WOC result for cluster 6 did not exhibit these variations, and appeared to be consistent with the other clusters.

\section{Results and Discussion} \label{sec:discuss}
\begin{figure}
\begin{center}
\includegraphics[width=0.55\columnwidth,trim={0 0.4cm 1.5cm 1.1cm},clip]{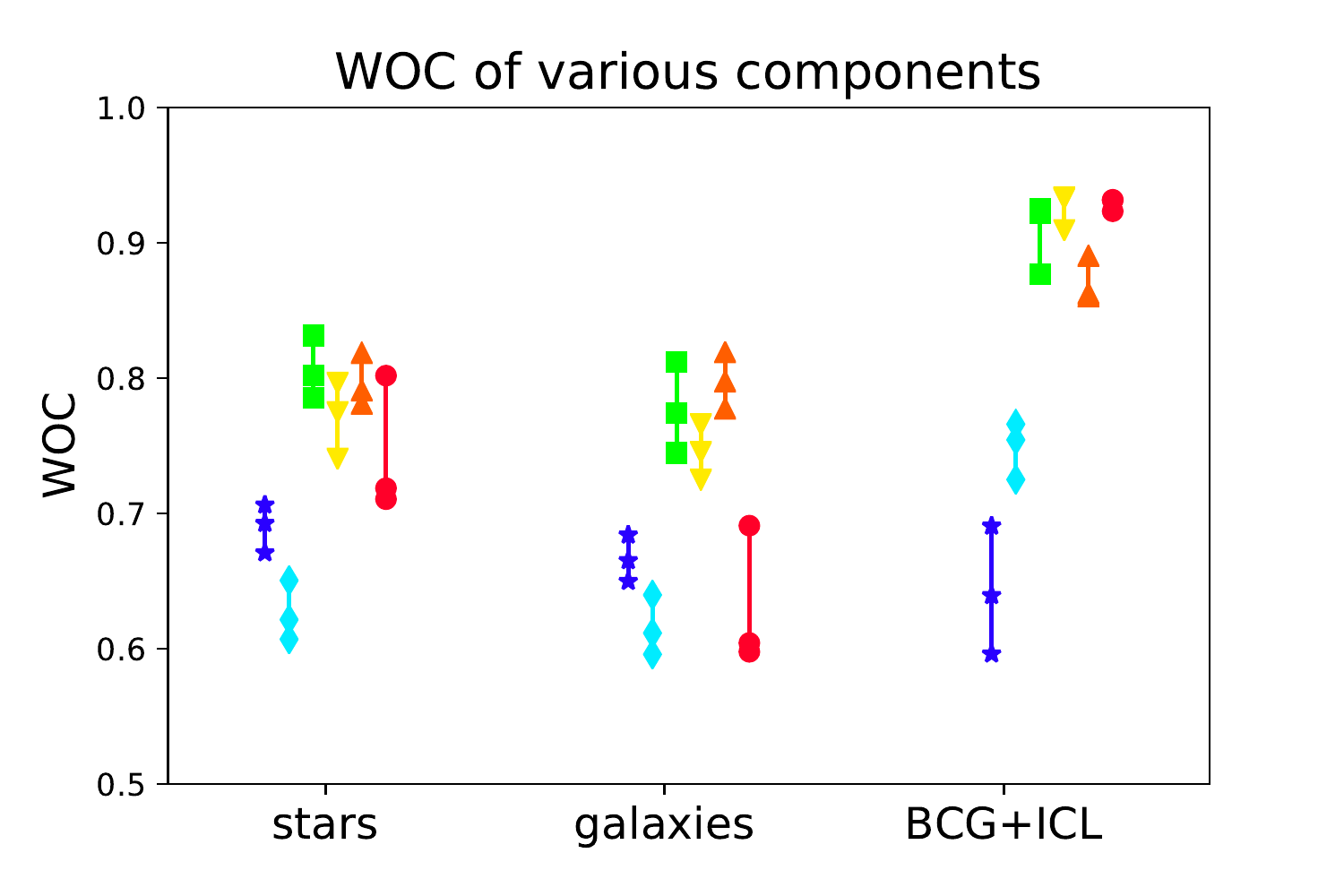}
\includegraphics[width=0.55\columnwidth,trim={0 0 1.5cm 1.1cm},clip]{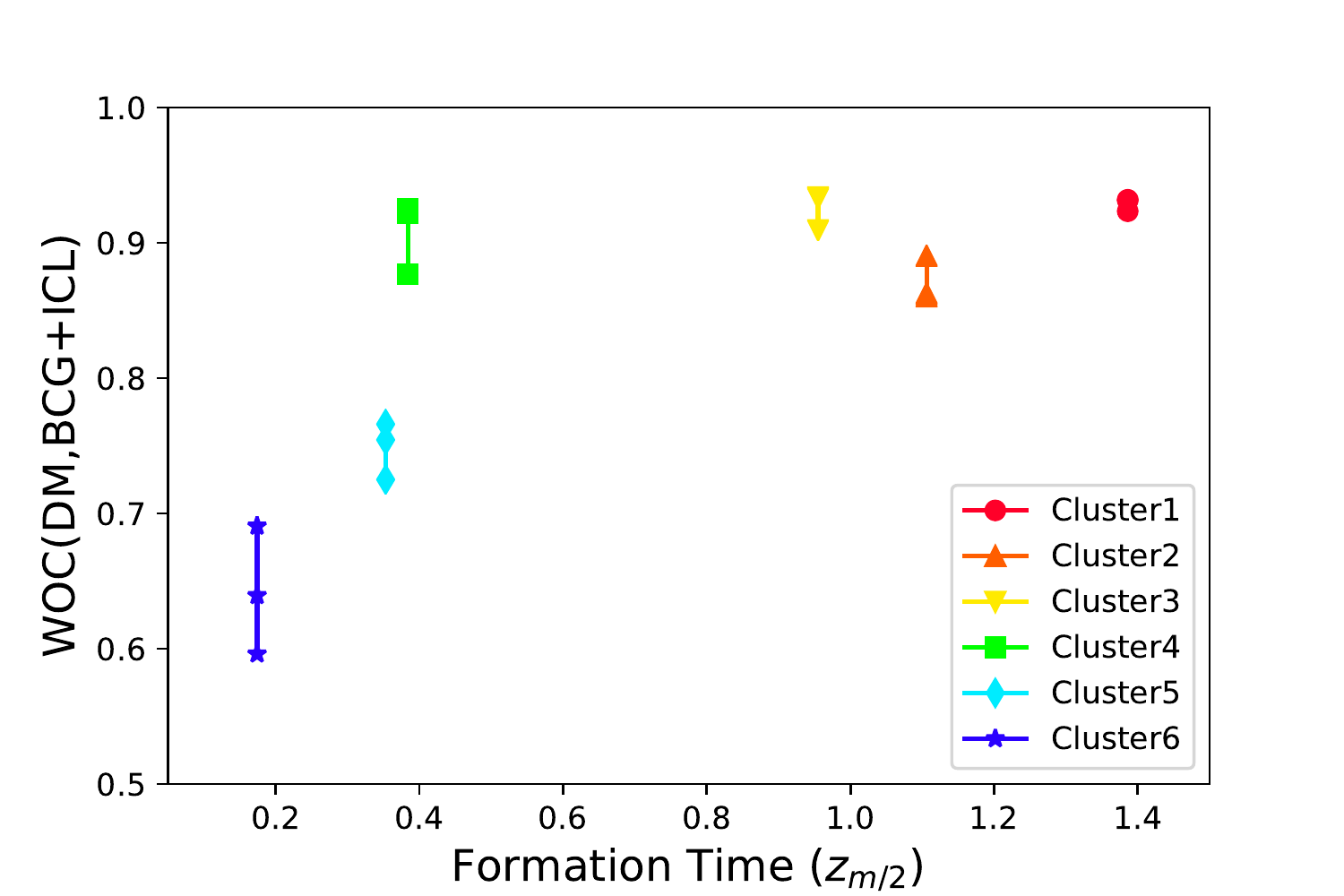}
\includegraphics[width=0.55\columnwidth,trim={0 0 1.5cm 1.1cm},clip]{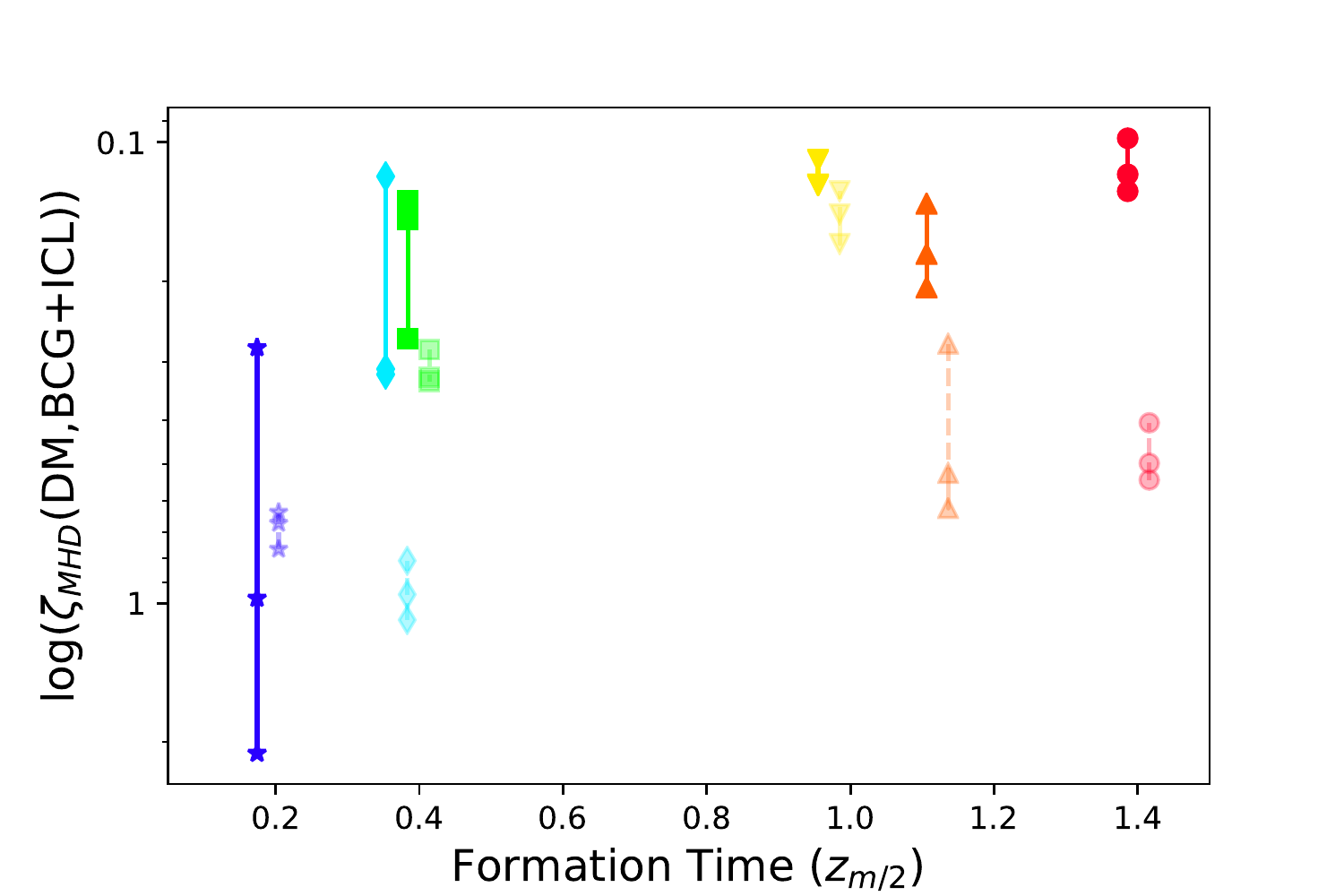}
\caption{WOC results for different components of galaxy clusters (upper panel), different dynamic stages of galaxy clusters (middle panel), and the MHD results for different dynamic stages of galaxy clusters (lower panel) are plotted.  A smoothing factor of 17 is used for all images. The three data points with the same symbols come from the three projection angles of each galaxy cluster, which are connected with lines. The MHD method results using only the largest contour at each level are plotted as solid lines, while the results using all contours are plotted as dashed faint lines.  We shift the MHD results with all contours +0.03 in the x-axis for readability. See more details in Section \ref{sec:discuss}.}
\label{fig:Dynamic}
\end{center}
\end{figure}

\textit{\bf Luminous tracer for dark matter}: We calculated the WOC of the dark matter and various components of each galaxy cluster to investigate which component had a spatial distribution most similar to the dark matter in the clusters. We prepared maps containing (1) all stellar particles, which includes BCG, other member galaxies, and ICL, (2) cluster galaxies, which includes BCG and other member galaxies, (3) BCG and ICL. Among the various components (stellar particles, galaxies, BCG+ICL), the BCG+ICL component traced the dark matter best, except for the case of cluster 6 (see the upper panel in Figure \ref{fig:Dynamic}). The three data points with the same symbols come from the three projection angles of each galaxy cluster, which are connected by solid bars. A possible explanation for the fact that the BCG+ICL component (without other member galaxies) traces the dark matter better rather than all stellar particles, could be that the shape of gravitational potential generated by dark matter is smoother than the sharp local minimum of potential produced by the stellar particles from other member galaxies. 

\textit{\bf Evolutionary
stage estimator for galaxy clusters}: The more relaxed galaxy clusters, which have had enough time to virialize, could have more aligned distributions between the dark matter and the other components of the cluster. If the BCG+ICL component of more relaxed galaxy clusters has a distribution more similar to the dark matter in them, compared to dynamically younger galaxy clusters, the similarity in spatial distribution between dark matter and BCG+ICL could be used as a dynamical stage indicator. Accordingly, we investigated the relation between the WOC(DM, BCG+ICL) result and one of the relaxation level proxy, formation time ($z_{m/2}$). Between the sample galaxy clusters, the relaxed clusters showed a stronger similarity in spatial distribution between the dark matter and BCG+ICL than the dynamically young clusters (see the middle panel in Figure \ref{fig:Dynamic}). The WOC - $z_{m/2}$ relation shows a considerable trend, although cluster 4 seems to be off from the trend. Among the clusters with high WOC (clusters 1, 2, 3 and 4), cluster 4 exhibits a slightly larger variation with viewing angle.

Care should be taken when comparing between the results of the MHD method and the WOC method, since their ranges and scaling are different. In Figure \ref{fig:param}, taking the linear scale of $\zeta_{MHD}$ seems like the most appropriate choice when comparing with the WOC results. Nevertheless, we took the logarithmic scale of $\zeta_{MHD}$ to allow for an easier qualitative visual comparison.
The MHD result shows similar trend with formation time but with a larger variation inferred from the viewing angle (see the solid line in lower panel in Figure \ref{fig:Dynamic}). Specially, cluster 6 showed significantly different $\zeta_{MHD}$(DM,BCG+ICL) results for different projection angles. Modifying the MHD method taking account of all contours in the calculation (dashed faint line) reduces the sensitivity to the viewing angle, however the resulting trend with dynamical stage is broken. The modified MHD improves in the merging cluster case by taking account of sub-structures, while in the relaxed cluster case, the resulting $\zeta_{MHD}$ becomes larger compared to the one-to-one contour comparison. 

The WOC method is a non-parametric comparison of two distributions, which is easy to use and does not require any fitting that may cause numerical error. It quantifies the similarity of two spatial distributions robustly, under choices of a smoothing factor, center, and binning (Figure \ref{fig:Smooth}). Moreover, the WOC method works well for disconnected contour cases when there are multiple significant sub-structures, such as in the Bullet cluster or Coma cluster. Also it does not require the computation of individual contours, and thus it suffers less bias even when working with masked maps. Furthermore, WOC(DM, BCG+ICL) could be used as the dynamical stage estimator of galaxy cluster, using the WOC(DM, BCG+ICL)-$z_{m/2}$ relation. Note that, the MHD method could consider the sub-structure in a way similar to the WOC method, if we did not choose the largest contour for each level but rather used all contours for the level (see the dashed faint line in the right panel in Figure \ref{fig:Smooth} and lower panel in Figure \ref{fig:Dynamic}), however the MHD method does not show a significant trend in the  $\zeta_{MHD}$(DM,BCG+ICL)-$z_{m/2}$ relation. 

\textit{\bf Application on observational data}:  In the WOC calculation, we require precise density maps to decide the levels and give weights using the density level at the contour. However, a computation of a density requires a robust measurement of the zero level, which is typically challenging for observations of the ICL. Therefore, when applying the WOC method to real data, we should compute a conservative detection limit of the image, and determine the binning range adequately. If the outermost bin (the largest contour) is securely within the area defined by the detection limit, the dimmer pixels outside the largest contour do not affect the WOC result. Even within the secure region, the ICL density level at each contour has an error from the background fluctuations. In the WOC weight system, all density level values are normalized (see the equation \ref{eq:weight}), thus the absolute accuracy of each density level is not so important, rather the relative density levels within the secure area are considered. The WOC$_M$ (introduced in Section \ref{sec:Character}) could be more problematic for observational data because the measurement of the total mass of dark matter and total luminosity of BGC+ICL requires the pixel values outside the largest contour. In either case, we should determine the detection limit boundary conservatively, to minimize this problem.

\textit{\bf Possible applications}: Aside from 2-dimensional spatial distributions, the WOC method could also be applicable to any two-parameter spaces, such as color-magnitude diagrams, phase-space diagrams, etc. One could also use our methodology to compare the characteristics of species in different galaxy clusters. Moreover, the WOC method could be simply extended to 3-dimensional spatial distributions by measuring `overlapping volume'.

In future work we will apply the method to galaxy clusters simulated in hydrodynamical simulation, such as Illustris TNG \citep{2019ComAC...6....2N} and Horizon Run5 \citep{2021ApJ...908...11L}, and study the relation between WOC(DM, BCG+ICL) and the cluster's dynamical stage (represented by virial ratio, sub-structure ratio, center of mass-BCG offset, formation time, etc.). 

Creating mock observation maps that represent different ICL formation origins (such as BCG major merger, tidal stripping or in-situ formation) and comparing them with observed ICL maps could constrain the origin of ICL.
Dynamically young clusters, whose ICL and dark matter are not well aligned, may possess a blue colored ICL, indicating freshly stripped stars.
Studying the ICL color and the WOC(DM, BCG+ICL) together can provide additional information, allowing us to test such hypotheses.

Furthermore, we could utilize the method to constrain dark matter models such as the self-interacting dark matter model (SIDM) or the cold dark matter model (CDM), which predict different tidal interaction histories \citep{2021arXiv210903257S}.
Because our methodology is trivially extendable to $N$-dimensions, we will search for use cases outside of astronomy that also require a spatial comparison of multiple distributions.

\section{Summary} \label{sec:conclusion}

We developed a new methodology to quantify the similarity of two or more 2-dimensional spatial distributions, which we call the Weighted Overlap Coefficient (WOC) method.
In this methodology, we calculate the fraction of overlapping area between two distributions at various density threshold levels, while weighting using the signal strength, and normalizing the similarity measure between 0 and 1. We compared this with one of the previous methods used in the literature, the Modified Hausdorff Distance (MHD) method. Our methodology is robust when choosing smoothing factor, center, and binning (Figure \ref{fig:Smooth}), and performs well even with the existence of multiple sub-structures, even those far away ($\sim$ 1 Mpc) from the center (Figure \ref{fig:process2}). Furthermore, the fixed range of the WOC result allows us to interpret the result easily and to compare between different cluster samples in a fair way. 

Applying the WOC method to six galaxy clusters simulated in GRT simulations, we found that the spatial distribution of the BCG+ICL component was a better match with dark matter than other cluster components, such as all stellar particles or all member galaxies (upper panel in Figure \ref{fig:Dynamic}). Most interestingly, galaxy clusters that were more relaxed (with earlier formation time, $z_{m/2}$) showed higher WOC(DM, BCG+ICL) results, which would enable us to use it as cluster's dynamical stage indicator (middle panel in Figure \ref{fig:Dynamic}). Conversely, we found that the MHD results show weaker trend with the dynamical state of the cluster (lower panel in Figure \ref{fig:Dynamic}). 

Our methodology may have broad application across astronomy and the physical sciences. 
 The software is available on GitHub\footnote{\texttt{WOC} codebase: \url{https://github.com/csabiu/WOC}.} under a GNU General Public License and version 0.0.1 used in this analysis is archived in Zenodo \citep{cristiano_sabiu_2022_6523107}.

\acknowledgments
We thank the anonymous referee for useful comments that have improved this paper.
This research was supported by the Korea Astronomy and Space Science Institute under the R\&D program(Project No. 2022-1-830-05) supervised by the Ministry of Science and ICT.
CGS acknowledges support via the Basic Science Research Program from the National Research Foundation of South Korea (NRF) funded by the Ministry of Education (2018R1A6A1A06024977 and 2020R1I1A1A01073494). KWC was supported by the NRF grant funded by the Korea government (MSIT) (2021R1F1A1045622). The simulations were supported by the National Supercomputing Center with supercomputing resources including technical support (KSC-2020-CRE-0297). H.S.H. acknowledges the support by the NRF grant funded by the MSIT (No. 2021R1A2C1094577). J.K. was supported by a KIAS individual grant (KG039603) via the Center for Advanced Computation at the Korea Institute for Advanced Study.

\vspace{5mm}

\software{WOC \citep{cristiano_sabiu_2022_6523107}, astropy \citep{2013A&A...558A..33A}
          }

\bibliography{sample63}{}

\begin{thebibliography}{}
\expandafter\ifx\csname natexlab\endcsname\relax\def\natexlab#1{#1}\fi
\providecommand{\url}[1]{\href{#1}{#1}}
\providecommand{\dodoi}[1]{doi:~\href{http://doi.org/#1}{\nolinkurl{#1}}}
\providecommand{\doeprint}[1]{\href{http://ascl.net/#1}{\nolinkurl{http://ascl.net/#1}}}
\providecommand{\doarXiv}[1]{\href{https://arxiv.org/abs/#1}{\nolinkurl{https://arxiv.org/abs/#1}}}

\bibitem[{{Alonso Asensio} {et~al.}(2020){Alonso Asensio}, {Dalla Vecchia},
  {Bah{\'e}}, {Barnes}, \& {Kay}}]{2020MNRAS.494.1859A}
{Alonso Asensio}, I., {Dalla Vecchia}, C., {Bah{\'e}}, Y.~M., {Barnes}, D.~J.,
  \& {Kay}, S.~T. 2020, \mnras, 494, 1859, \dodoi{10.1093/mnras/staa861}

\bibitem[{{Astropy Collaboration} {et~al.}(2013){Astropy Collaboration},
  {Robitaille}, {Tollerud}, {Greenfield}, {Droettboom}, {Bray}, {Aldcroft},
  {Davis}, {Ginsburg}, {Price-Whelan}, {Kerzendorf}, {Conley}, {Crighton},
  {Barbary}, {Muna}, {Ferguson}, {Grollier}, {Parikh}, {Nair}, {Unther},
  {Deil}, {Woillez}, {Conseil}, {Kramer}, {Turner}, {Singer}, {Fox}, {Weaver},
  {Zabalza}, {Edwards}, {Azalee Bostroem}, {Burke}, {Casey}, {Crawford},
  {Dencheva}, {Ely}, {Jenness}, {Labrie}, {Lim}, {Pierfederici}, {Pontzen},
  {Ptak}, {Refsdal}, {Servillat}, \& {Streicher}}]{2013A&A...558A..33A}
{Astropy Collaboration}, {Robitaille}, T.~P., {Tollerud}, E.~J., {et~al.} 2013,
  \aap, 558, A33, \dodoi{10.1051/0004-6361/201322068}

\bibitem[{{Behroozi} {et~al.}(2013){Behroozi}, {Wechsler}, \&
  {Wu}}]{2013ApJ...762..109B}
{Behroozi}, P.~S., {Wechsler}, R.~H., \& {Wu}, H.-Y. 2013, \apj, 762, 109,
  \dodoi{10.1088/0004-637X/762/2/109}

\bibitem[{{Chun} {et~al.}(2022){Chun}, {Shin}, {Smith}, {Ko}, \&
  {Yoo}}]{2022ApJ...925..103C}
{Chun}, K., {Shin}, J., {Smith}, R., {Ko}, J., \& {Yoo}, J. 2022, \apj, 925,
  103, \dodoi{10.3847/1538-4357/ac2cbe}

\bibitem[{{Contini} {et~al.}(2014){Contini}, {De Lucia}, {Villalobos}, \&
  {Borgani}}]{2014MNRAS.437.3787C}
{Contini}, E., {De Lucia}, G., {Villalobos}, {\'A}., \& {Borgani}, S. 2014,
  \mnras, 437, 3787, \dodoi{10.1093/mnras/stt2174}

\bibitem[{{Contini} {et~al.}(2019){Contini}, {Yi}, \&
  {Kang}}]{2019ApJ...871...24C}
{Contini}, E., {Yi}, S.~K., \& {Kang}, X. 2019, \apj, 871, 24,
  \dodoi{10.3847/1538-4357/aaf41f}

\bibitem[{{DeMaio} {et~al.}(2018){DeMaio}, {Gonzalez}, {Zabludoff}, {Zaritsky},
  {Connor}, {Donahue}, \& {Mulchaey}}]{2018MNRAS.474.3009D}
{DeMaio}, T., {Gonzalez}, A.~H., {Zabludoff}, A., {et~al.} 2018, \mnras, 474,
  3009, \dodoi{10.1093/mnras/stx2946}

\bibitem[{{Dubuisson} \& {Jain}(1994)}]{576361}
{Dubuisson}, M.~., \& {Jain}, A.~K. 1994, in Proceedings of 12th International
  Conference on Pattern Recognition, Vol.~1, 566--568 vol.1,
  \dodoi{10.1109/ICPR.1994.576361}

\bibitem[{{Dutton}(2009)}]{2009MNRAS.396..121D}
{Dutton}, A.~A. 2009, \mnras, 396, 121,
  \dodoi{10.1111/j.1365-2966.2009.14741.x}

\bibitem[{{Feldmeier} {et~al.}(2002){Feldmeier}, {Mihos}, {Morrison}, {Rodney},
  \& {Harding}}]{2002ApJ...575..779F}
{Feldmeier}, J.~J., {Mihos}, J.~C., {Morrison}, H.~L., {Rodney}, S.~A., \&
  {Harding}, P. 2002, \apj, 575, 779, \dodoi{10.1086/341472}

\bibitem[{{Fujita} {et~al.}(2018){Fujita}, {Umetsu}, {Ettori}, {Rasia},
  {Okabe}, \& {Meneghetti}}]{2018ApJ...863...37F}
{Fujita}, Y., {Umetsu}, K., {Ettori}, S., {et~al.} 2018, \apj, 863, 37,
  \dodoi{10.3847/1538-4357/aacf05}

\bibitem[{{Furnell} {et~al.}(2021){Furnell}, {Collins}, {Kelvin}, {Baldry},
  {James}, {Manolopoulou}, {Mann}, {Giles}, {Bermeo}, {Hilton}, {Wilkinson},
  {Romer}, {Vergara}, {Bhargava}, {Stott}, {Mayers}, \&
  {Viana}}]{2021MNRAS.502.2419F}
{Furnell}, K.~E., {Collins}, C.~A., {Kelvin}, L.~S., {et~al.} 2021, \mnras,
  502, 2419, \dodoi{10.1093/mnras/stab065}

\bibitem[{{Gonzalez} {et~al.}(2005){Gonzalez}, {Zabludoff}, \&
  {Zaritsky}}]{2005ApJ...618..195G}
{Gonzalez}, A.~H., {Zabludoff}, A.~I., \& {Zaritsky}, D. 2005, \apj, 618, 195,
  \dodoi{10.1086/425896}

\bibitem[{{Hwang} {et~al.}(2014){Hwang}, {Geller}, {Diaferio}, {Rines}, \&
  {Zahid}}]{2014ApJ...797..106H}
{Hwang}, H.~S., {Geller}, M.~J., {Diaferio}, A., {Rines}, K.~J., \& {Zahid},
  H.~J. 2014, \apj, 797, 106, \dodoi{10.1088/0004-637X/797/2/106}

\bibitem[{{Jim{\'e}nez-Teja} {et~al.}(2019){Jim{\'e}nez-Teja}, {Dupke}, {Lopes
  de Oliveira}, {Xavier}, {Coelho}, {Chies-Santos}, {L{\'o}pez-Sanjuan},
  {Alvarez-Candal}, {Costa-Duarte}, {Telles}, {Hernandez-Jimenez},
  {Ben{\'\i}tez}, {Alcaniz}, {Cenarro}, {Crist{\'o}bal-Hornillos},
  {Ederoclite}, {Mar{\'\i}n-Franch}, {Mendes de Oliveira}, {Moles},
  {Sodr{\'e}}, {Varela}, \& {V{\'a}zquez Rami{\'o}}}]{2019A&A...622A.183J}
{Jim{\'e}nez-Teja}, Y., {Dupke}, R.~A., {Lopes de Oliveira}, R., {et~al.} 2019,
  \aap, 622, A183, \dodoi{10.1051/0004-6361/201833547}

\bibitem[{{Kantorovich}(1960)}]{Kantorovich}
{Kantorovich}, L.~V. 1960, Management Science, 6, 366,
  \dodoi{10.1287/mnsc.6.4.366}

\bibitem[{{Kluge} {et~al.}(2020){Kluge}, {Neureiter}, {Riffeser}, {Bender},
  {Goessl}, {Hopp}, {Schmidt}, {Ries}, \& {Brosch}}]{2020ApJS..247...43K}
{Kluge}, M., {Neureiter}, B., {Riffeser}, A., {et~al.} 2020, \apjs, 247, 43,
  \dodoi{10.3847/1538-4365/ab733b}

\bibitem[{{Ko} \& {Jee}(2018)}]{2018ApJ...862...95K}
{Ko}, J., \& {Jee}, M.~J. 2018, \apj, 862, 95, \dodoi{10.3847/1538-4357/aacbda}

\bibitem[{{Lee} {et~al.}(2021){Lee}, {Shin}, {Snaith}, {Kim}, {Few},
  {Devriendt}, {Dubois}, {Cox}, {Hong}, {Kwon}, {Park}, {Pichon}, {Kim},
  {Gibson}, \& {Park}}]{2021ApJ...908...11L}
{Lee}, J., {Shin}, J., {Snaith}, O.~N., {et~al.} 2021, \apj, 908, 11,
  \dodoi{10.3847/1538-4357/abd08b}

\bibitem[{{Lin} \& {Mohr}(2004)}]{2004ApJ...617..879L}
{Lin}, Y.-T., \& {Mohr}, J.~J. 2004, \apj, 617, 879, \dodoi{10.1086/425412}

\bibitem[{{Mihos}(2016)}]{2016IAUS..317...27M}
{Mihos}, J.~C. 2016, in The General Assembly of Galaxy Halos: Structure, Origin
  and Evolution, ed. A.~{Bragaglia}, M.~{Arnaboldi}, M.~{Rejkuba}, \&
  D.~{Romano}, Vol. 317, 27--34, \dodoi{10.1017/S1743921315006857}

\bibitem[{{Mihos} {et~al.}(2005){Mihos}, {Harding}, {Feldmeier}, \&
  {Morrison}}]{2005ApJ...631L..41M}
{Mihos}, J.~C., {Harding}, P., {Feldmeier}, J., \& {Morrison}, H. 2005, \apjl,
  631, L41, \dodoi{10.1086/497030}

\bibitem[{{Mihos} {et~al.}(2017){Mihos}, {Harding}, {Feldmeier}, {Rudick},
  {Janowiecki}, {Morrison}, {Slater}, \& {Watkins}}]{2017ApJ...834...16M}
{Mihos}, J.~C., {Harding}, P., {Feldmeier}, J.~J., {et~al.} 2017, \apj, 834,
  16, \dodoi{10.3847/1538-4357/834/1/16}

\bibitem[{{Montes} \& {Trujillo}(2019)}]{2019MNRAS.482.2838M}
{Montes}, M., \& {Trujillo}, I. 2019, \mnras, 482, 2838,
  \dodoi{10.1093/mnras/sty2858}

\bibitem[{{More} {et~al.}(2015){More}, {Diemer}, \&
  {Kravtsov}}]{2015ApJ...810...36M}
{More}, S., {Diemer}, B., \& {Kravtsov}, A.~V. 2015, \apj, 810, 36,
  \dodoi{10.1088/0004-637X/810/1/36}

\bibitem[{{Nelson} {et~al.}(2018){Nelson}, {Pillepich}, {Springel},
  {Weinberger}, {Hernquist}, {Pakmor}, {Genel}, {Torrey}, {Vogelsberger},
  {Kauffmann}, {Marinacci}, \& {Naiman}}]{2018MNRAS.475..624N}
{Nelson}, D., {Pillepich}, A., {Springel}, V., {et~al.} 2018, \mnras, 475, 624,
  \dodoi{10.1093/mnras/stx3040}

\bibitem[{{Nelson} {et~al.}(2019){Nelson}, {Springel}, {Pillepich},
  {Rodriguez-Gomez}, {Torrey}, {Genel}, {Vogelsberger}, {Pakmor}, {Marinacci},
  {Weinberger}, {Kelley}, {Lovell}, {Diemer}, \&
  {Hernquist}}]{2019ComAC...6....2N}
{Nelson}, D., {Springel}, V., {Pillepich}, A., {et~al.} 2019, Computational
  Astrophysics and Cosmology, 6, 2, \dodoi{10.1186/s40668-019-0028-x}

\bibitem[{Sabiu \& Yoo(2022)}]{cristiano_sabiu_2022_6523107}
Sabiu, C., \& Yoo, J. 2022, WOC weighted overlap coefficient, 0.0.1,  Zenodo,
  \dodoi{10.5281/zenodo.6523107}

\bibitem[{{Sampaio-Santos} {et~al.}(2021){Sampaio-Santos}, {Zhang}, {Ogando},
  {Shin}, {Golden-Marx}, {Yanny}, {Herner}, {Hilton}, {Choi}, {Gatti}, {Gruen},
  {Hoyle}, {Rau}, {De Vicente}, {Zuntz}, {Abbott}, {Aguena}, {Allam}, {Annis},
  {Avila}, {Bertin}, {Brooks}, {Burke}, {Carrasco Kind}, {Carretero}, {Chang},
  {Costanzi}, {da Costa}, {Diehl}, {Doel}, {Everett}, {Evrard}, {Flaugher},
  {Fosalba}, {Frieman}, {Garc{\'\i}a-Bellido}, {Gaztanaga}, {Gerdes},
  {Gruendl}, {Gschwend}, {Gutierrez}, {Hinton}, {Hollowood}, {Honscheid},
  {James}, {Jarvis}, {Jeltema}, {Kuehn}, {Kuropatkin}, {Lahav}, {Maia},
  {March}, {Marshall}, {Miquel}, {Palmese}, {Paz-Chinch{\'o}n}, {Plazas},
  {Sanchez}, {Santiago}, {Scarpine}, {Schubnell}, {Smith}, {Suchyta}, {Tarle},
  {Tucker}, {Varga}, \& {Wechsler}}]{2021MNRAS.501.1300S}
{Sampaio-Santos}, H., {Zhang}, Y., {Ogando}, R.~L.~C., {et~al.} 2021, \mnras,
  501, 1300, \dodoi{10.1093/mnras/staa3680}

\bibitem[{Schoener(1968)}]{schoener1968anolis}
Schoener, T.~W. 1968, Ecology, 49, 704

\bibitem[{{Sirks} {et~al.}(2021){Sirks}, {Oman}, {Robertson}, {Massey}, \&
  {Frenk}}]{2021arXiv210903257S}
{Sirks}, E.~L., {Oman}, K.~A., {Robertson}, A., {Massey}, R., \& {Frenk}, C.
  2021, arXiv e-prints, arXiv:2109.03257.
\newblock \doarXiv{2109.03257}

\bibitem[{{Springel}(2005)}]{2005MNRAS.364.1105S}
{Springel}, V. 2005, \mnras, 364, 1105,
  \dodoi{10.1111/j.1365-2966.2005.09655.x}

\bibitem[{Warren {et~al.}(2008)Warren, Glor, \&
  Turelli}]{warren2008environmental}
Warren, D.~L., Glor, R.~E., \& Turelli, M. 2008, Evolution: International
  Journal of Organic Evolution, 62, 2868

\bibitem[{{Yoo} {et~al.}(2021){Yoo}, {Ko}, {Kim}, \&
  {Kim}}]{2021MNRAS.508.2634Y}
{Yoo}, J., {Ko}, J., {Kim}, J.-W., \& {Kim}, H. 2021, \mnras, 508, 2634,
  \dodoi{10.1093/mnras/stab2707}

\bibitem[{{Zibetti} {et~al.}(2005){Zibetti}, {White}, {Schneider}, \&
  {Brinkmann}}]{2005MNRAS.358..949Z}
{Zibetti}, S., {White}, S. D.~M., {Schneider}, D.~P., \& {Brinkmann}, J. 2005,
  \mnras, 358, 949, \dodoi{10.1111/j.1365-2966.2005.08817.x}

\end{thebibliography}
\bibliographystyle{aasjournal}

\end{document}